\documentclass[sigconf]{acmart}

\usepackage{amsmath}
\usepackage{multirow}
\usepackage{bbm}
\usepackage{caption}
\usepackage{balance}

\AtBeginDocument{%
  \providecommand\BibTeX{{%
    \normalfont B\kern-0.5em{\scshape i\kern-0.25em b}\kern-0.8em\TeX}}}

\copyrightyear{2024}
\acmYear{2024}
\setcopyright{rightsretained}
\acmConference[MMASIA '24]{ACM Multimedia Asia}{December 3--6,
2024}{Auckland, New Zealand}
\acmBooktitle{ACM Multimedia Asia (MMASIA '24), December 3--6, 2024,
Auckland, New Zealand}
\acmDOI{10.1145/3696409.3700225}
\acmISBN{979-8-4007-1273-9/24/12}

\makeatletter
\gdef\@copyrightpermission{
  \begin{minipage}{0.3\columnwidth}
   \href{https://creativecommons.org/licenses/by/4.0/}{\includegraphics[width=0.90\textwidth]{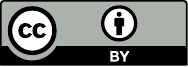}}
  \end{minipage}\hfill
  \begin{minipage}{0.7\columnwidth}
   \href{https://creativecommons.org/licenses/by/4.0/}{This work is licensed under a Creative Commons Attribution International 4.0 License.}
  \end{minipage}
  \vspace{5pt}
}
\makeatother

\begin{document}

\title[Enhancing Modality Representation and Alignment for Multimodal Cold-start AL]{Enhancing Modality Representation and Alignment for Multimodal Cold-start Active Learning}


\author{Meng Shen}
\orcid{0000-0003-2502-3500}
\affiliation{%
  \institution{Nanyang Technological University}
  \city{Singapore}
  \country{SG}
  }
\email{meng005@e.ntu.edu.sg}

\author{Yake Wei}
\orcid{0009-0008-8779-0637}
\affiliation{%
  \institution{Renmin University of China}
  \city{Beijing}
  \country{CN}
  }
\email{yakewei@ruc.edu.cn}

\author{Jianxiong Yin}
\orcid{0000-0003-4686-2768}
\affiliation{%
  \institution{NVIDIA AI Tech Center}
  \city{Singapore}
  \country{SG}
  }
\email{jianxiongy@nvidia.com}

\author{Deepu Rajan}
\orcid{0000-0001-7788-8368}
\affiliation{%
  \institution{Nanyang Technological University}
  \city{Singapore}
  \country{SG}
  }
\email{asdrajan@ntu.edu.sg}

\author{Di Hu}
\orcid{0000-0002-7118-6733}
\affiliation{%
  \institution{Renmin University of China}
  \city{Beijing}
  \country{CN}
  }
\email{dihu@ruc.edu.cn}

\author{Simon See}
\orcid{0000-0002-4958-9237}
\affiliation{%
  \institution{NVIDIA AI Tech Center}
  \city{Singapore}
  \country{SG}
  }
\email{ssee@nvidia.com}

\renewcommand{\shortauthors}{Meng Shen et al.}
\begin{abstract}
  Training multimodal models requires a large amount of labeled data. Active learning (AL) aim to reduce labeling costs. Most AL methods employ warm-start approaches, which rely on sufficient labeled data to train a well-calibrated model that can assess the uncertainty and diversity of unlabeled data. However, when assembling a dataset, labeled data are often scarce initially, leading to a cold-start problem. Additionally, most AL methods seldom address multimodal data, highlighting a research gap in this field. Our research addresses these issues by developing a two-stage method for \textbf{M}ulti-\textbf{M}odal \textbf{C}old-\textbf{S}tart \textbf{A}ctive \textbf{L}earning (\textbf{MMCSAL}). 
  Firstly, we observe the modality gap, a significant distance between the centroids of representations from different modalities, when only using cross-modal pairing information as self-supervision signals. This modality gap affects data selection process, as we calculate both uni-modal and cross-modal distances. To address this, we introduce uni-modal prototypes to bridge the modality gap. Secondly, conventional AL methods often falter in multimodal scenarios where alignment between modalities is overlooked. Therefore, we propose enhancing cross-modal alignment through regularization, thereby improving the quality of selected multimodal data pairs in AL. Finally, our experiments demonstrate MMCSAL's efficacy in selecting multimodal data pairs across three multimodal datasets.
\end{abstract}




\maketitle

\section{Introduction}

\begin{figure}[t]
  \centering
   \includegraphics[width=0.9\linewidth]{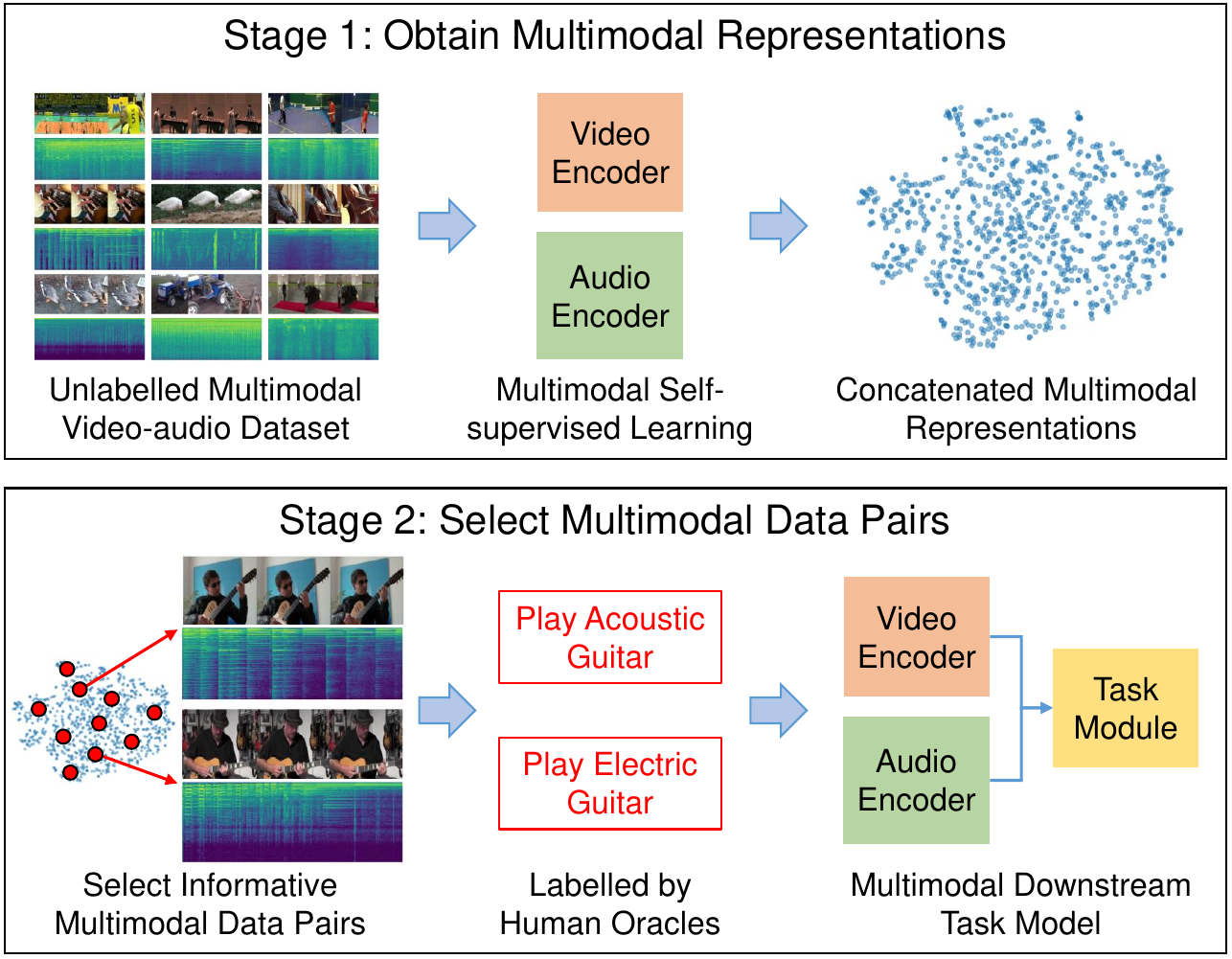}
   \caption{MMCSAL Framework (using video-audio data as an example): In Stage 1, MMSSL is applied to a large unlabeled dataset to derive concatenated multimodal representations from unimodal features. In Stage 2, a data selection strategy samples informative multimodal data pairs, which are then annotated by human oracles. The labeled samples are subsequently used to train a downstream task model.}
   \label{fig:active_learning}
\end{figure}

Deep multimodal networks process multimodal data pairs consisting of various modalities, harnessing the complementary information across these modalities \cite{DBLP:journals/jstsp/ZhangYHD20, DBLP:journals/pami/BaltrusaitisAM19}. However, their effectiveness hinges on the availability of a substantial volume of labeled multimodal data for training. To alleviate the issue of expensive data labeling, active learning (AL) strategies are widely applied to selectively annotate the most significant data samples \cite{DBLP:journals/csur/RenXCHLGCW22}. Most AL strategies initiate with a randomly selected subset of data, obtaining label annotations for these samples to start the data evaluation and selection, a process known as \textbf{Warm-start}. Nevertheless, in practice, assembling a multimodal dataset often starts with limited or even no labels. This poses a challenge for uncertainty-based, diversity-based, and hybrid AL strategies. These strategies rely on initial labels to construct a model capable of assessing the uncertainty and diversity of the data. A lack of sufficient initial labels leads to the \textbf{Cold-start} problem in AL, where there is inadequate information to initiate the active learning loop effectively \cite{cold_start_al_nlp, chen2024making}.

Researchers have explored to address the cold-start problem in uni-modal active learning, introducing a two-stage framework incorporating self-supervised learning (SSL) with data selection \cite{cold_start_al_nlp, chen2024making, al_on_a_budget, pt4al}. However, there exists specific problems in multimodal scenarios expected to be resolved. In \textbf{Stage 1}, multi-modal self-supervised learning (MMSSL) typically leverages inherent cross-modal pairing information to guide model training without labels. However, we observe that sole reliance on cross-modal information in MMSSL creates a modality gap, where the centroid of each uni-modal representations is far away from another. This modality gap can skew the distance measurements among multimodal data pairs, affecting the data selection process. 

In \textbf{Stage 2}, after completing MMSSL to generate semantically rich data representations, the initial batch of data samples is selected for annotation by human oracles. These newly labeled samples are then used to train downstream task models. Existing methods, however, predominantly focus on uni-modal data, such as text sentences \cite{cold_start_al_nlp} or images \cite{chen2024making, al_on_a_budget, pt4al, activeft}, and may not effectively handle multimodal data pairs. To the best of our knowledge, current cold-start AL approaches do not account for modality alignment in the selection of multimodal data pairs. This oversight neglects the potential for harnessing more complementary information, which could enhance the performance of multimodal downstream tasks.

To overcome the limitation in \textbf{Stage 1}, we propose enhancing MMSSL for AL, aimed at bridging the modality gap and improving distance estimation for multimodal data pairs. Our approach integrates uni-modal prototypes into cross-modal contrastive learning, yielding more balanced representations for both uni-modal and cross-modal feature densities and narrowing the modality gap. In \textbf{Stage 2}, addressing cross-modal alignment, which has not been previously explored in AL, is crucial for selecting multimodal data pairs. We introduce a regularization term dedicated to optimizing the modality alignment of the selected data subset. \textbf{Figure \ref{fig:active_learning}} illustrates the two stages for addressing the cold-start problem in multimodal AL. Empirical findings suggest that enhancing cross-modal similarity during data selection benefits the training of downstream multimodal tasks by leveraging more complementary information. Consequently, we develop a novel method, \textbf{M}ulti\textbf{M}odal \textbf{C}old-\textbf{S}tart \textbf{A}ctive \textbf{L}earning (\textbf{MMCSAL}), which strategically addresses these challenges. We summarize our contributions as followings:
\begin{itemize}
    \item We introduce a MMSSL method that incorporates cross-modal contrastive learning and uni-modal prototypical learning, which aims to reduce the modality gap and provide more accurate distance estimation for multimodal data selection.
    \item We propose a novel multimodal data selection strategy that improves the modality alignment of the selected data subset, leading to improved performance for downstream tasks.
    \item Through empirical evaluation, our method shows enhanced ability in selecting multimodal data pairs for cold-start multimodal AL across three multimodal classification datasets: Food101, KineticsSound and VGGSound, covering textual, auditory and visual modalities.
\end{itemize}

\section{Related Works}
\label{sec.2}

\textbf{Multimodal Self-supervised Learning} constructs semantically rich multimodal representations without the need for human annotations. These methods can be categorized by their learning objectives into contrastive discrimination, clustering, and masked token prediction \cite{mmssl_survey}. \textbf{Contrastive} methods treat paired multimodal samples as positive pairs and unpaired ones as negative pairs, aiming to minimize the representation distance between positive pairs while maximizing it for negative pairs during training. Models trained with contrastive learning, such as CLIP \cite{clip}, an image-text model, have produced image and text representations with aligned cross-modal semantic information, demonstrating robust zero-shot capabilities. Furthermore, contrastive learning has been successfully applied to various modality combinations, including video-text \cite{video_clip, crossclr}, audio-text \cite{clap}, and video-audio \cite{DBLP:conf/iclr/MaZMS21,DBLP:conf/aaai/Jenni0C23}. \textbf{Clustering} aims to group data with similar semantic information. In the realm of visual SSL, SwAV \cite{DBLP:conf/nips/CaronMMGBJ20} performs online clustering, assigning multiple instances to  prototypes and learning representations by predicting the cluster assignment of one view from another view. PCL \cite{prototypical}  combines contrastive learning for extracting useful visual features with cluster assignment prediction to enhance the semantic structure of the representations. In the multimodal domain, XDC \cite{DBLP:conf/nips/Alwassel0KTGT20} uses the clustering assignments of one modality as prediction targets for another, thereby deriving multimodal representations. AV-HuBert \cite{DBLP:conf/iclr/ShiHLM22} performs clustering on unmasked visual and auditory sequences, forcing the masked visual and auditory tokens to predict the clustering results. 

\begin{figure*}[ht!]
  \centering
   \includegraphics[width=0.98\linewidth]{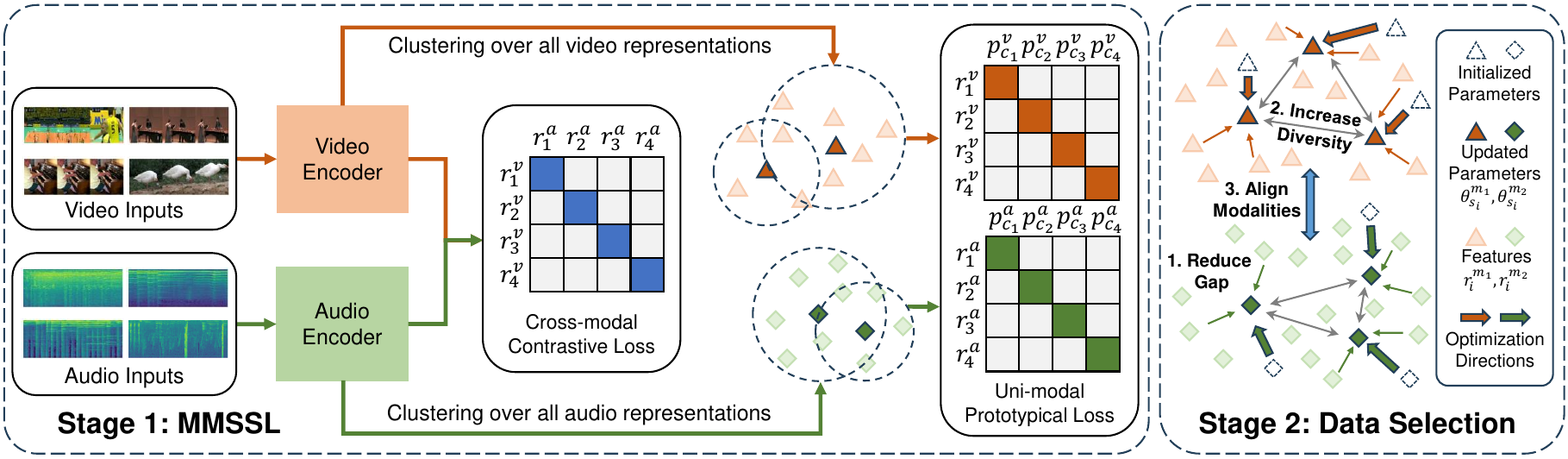}
   \caption{Our method (use audio/video as an example): In Stage 1, we employ uni-modal prototypical loss and cross-modal contrastive loss; In Stage 2, our selection reduces distribution gap while maintaining diversity and modality alignment.}
   \label{fig:mmssl}
\end{figure*}

\noindent \textbf{Cold-start Active Learning} aims to select the initial batch of data samples in circumstances where no labels are available. An effective cold-start AL strategy should identify a subset of samples that are more informative and diverse compared to a randomly chosen set. Due to the absence  of labels, SSL has been widely adopted to mitigate the cold-start problem in active learning. ALPS \cite{cold_start_al_nlp} utilizes self-supervised masked language modeling based on BERT \cite{DBLP:conf/naacl/DevlinCLT19} to create surprisal embeddings for each sentence sample, identifying more significant samples. PT4AL \cite{pt4al} performs a pretext task before data selection and identifies significant positive correlations between rotation prediction and image classification losses, and between colorization and segmentation losses. It ranks samples based on their pretext task loss, selecting the most challenging ones for the initial cold-start batch. CSVAL \cite{chen2024making} uses MoCo-v2 \cite{mocov2} as a SSL strategy to obtain image representations, then generates pseudo labels through K-means to train a classification model. It adopts Dataset Map \cite{dataset-map} to select out easy-to-learn samples to form the first batch. These samples are also found to be hard-to-contrast samples as they are hard to be distinguished from others, indicating their typicality. ActiveFT \cite{activeft} proposes to utilize data representations provided by pretrained models such as DINO \cite{DBLP:conf/iccv/CaronTMJMBJ21} to select a subset of data for label annotations and subsequent model fine-tuning. It minimizes the data distribution gap between the selected subset and the entire unlabeled dataset and regularizes the diversity to keep data selection from collapsing to centroid points.

These cold-start AL methods are devised for unimodal tasks. In this work, we aim to address the cold-start problem for multimodal AL by designing a suitable MMSSL to improve representation quality and enhancing modality alignment during data selection.

\section{Proposed Method}

\subsection{Multimodal Cold-start AL Framework}
We illustrate the multimodal cold-start active learning framework in \textbf{Figure \ref{fig:active_learning}} and provide details about our method in \textbf{Figure \ref{fig:mmssl}}. In the first stage, we are provided with an unlabeled multimodal dataset $X^U = \{(x^{m_1}_i, x^{m_2}_i)\}_{i\in [N]}$, consisting of $N$ data pairs across modalities $m_1$ and $m_2$. MMSSL is used to train a two-stream multimodal feature encoder, $f^{m_1}(x^{m_1}_i)$ and $f^{m_2}(x^{m_2}_i)$, for each modality, mapping data samples to representations $r^{m_1}$ and $r^{m_2}$ in the normalized high-dimensional feature space $\mathbbm{R}^C$. The concatenated multimodal representations, $\langle r^{m_1}_i \oplus r^{m_2}_i \rangle$, encapsulate the combined characteristics of both modalities. 

In the second stage, we allocate a label budget $B$, typically ranging from 1\% to 10\% of the total dataset size $N$, for acquiring labels from human oracles. Using the multimodal representations $R^U = \{\langle r^{m_1}_i \oplus r^{m_2}_i \rangle\}_{i\in [N]}$ from the first stage, we apply an AL strategy $S_{AL}$ to select a subset $X^S = \{(x^{m_1}_j, x^{m_2}_j)\}_{j\in[B]}$ from the entire unlabeled dataset $X^U$. This budget $B$ is used to assign labels $\{y_j\}_{j\in[B]}$ to the selected subset, forming a labeled subset $X^L= \{(x^{m_1}_j, x^{m_2}_j, y_j)\}_{j\in[B]}$. The labeled subset is then used to train a downstream task model.

\subsection{Multimodal Contrastive Learning with Uni-modal Prototypes}
\label{sec:mmssl}
The objective of MMSSL is to produce high-quality representations for multimodal data that can accurately capture both the uni- and cross-modal relationships. To first learn the cross-modal connections, we can seamlessly adopt cross-modal contrastive learning methods such as CLIP \cite{clip}. The idea is to bring the representations of two modalities from the same pair closer and push away those from the unpaired data. Specifically, for a batch of $K$ data pairs, the process involves maximizing the cosine similarity between L2-normalized modality representations for the $K$ paired samples and minimizing it for the $K\times(K - 1)$ unpaired modality combinations. We optimize the InfoNCE loss \cite{infonce}, with the temperature parameter $\tau$ set to 0.07, as recommended by studies such as \cite{DBLP:conf/cvpr/He0WXG20, mocov2}:

\begin{equation}
\label{eq:cross_modal_loss}
\begin{aligned}
    \mathcal{L}_{\text{Cross}}=-\underset{i\in [K]}{\mathbb{E}}  \left[ \frac{1}{2} \log  \frac{\exp({\cos(r^{m_1}_i, r^{m_2}_i) /\tau})}{ \sum_{j\in [K]} \exp({\cos(r^{m_1}_i, r^{m_2}_j) /\tau}) } + \right. \\
     \left. \frac{1}{2} \log  \frac{\exp({\cos(r^{m_2}_i, r^{m_1}_i) /\tau})}{ \sum_{j\in [K]} \exp({\cos(r^{m_2}_i, r^{m_1}_j) /\tau}) }\right].
\end{aligned}
\end{equation}

However, measuring the distance between multimodal sample pairs based solely on cross-modal distance may not be sufficiently accurate due to the modality domain gap. As shown in \cite{ModalityGap}, there is a modality gap in representations trained by contrastive loss, affecting the accuracy of distance measurement between samples which is crucial in our data selection. This modality gap phenomenon is also observed in our experiments across three datasets. As detailed in \textbf{Table \ref{table0:modality_gap}}, the Euclidean distance between the centroids of two modality representations $\Delta_{\text{gap}}$ is larger without uni-modal prototypes. To reduce this modality gap, we incorporate uni-modal prototypes that reshape the multimodal representations by optimizing the uni-modal prototypical loss within each modality.  

We construct uni-modal prototypes individually for each modality and visualize our MMSSL method in the left side of \textbf{Figure \ref{fig:mmssl}}. Here, we follow the implementation from PCL \cite{prototypical}. Take the first modality $m_1$ as an example,  we perform K-means clustering on its uni-modal representations $ \{ r^{m_1}_i \}_{i\in [N]}$ to produce $C$ clusters, and the centroid of each cluster is referred to as a prototype $p^{m_1}$. The same process is performed for the other modality individually. We assign the representation $r^{m_1}_i$ to the prototype $p^{m_1}_{c_i}$ of its cluster $c_i$ as a positive pair. Conversely, we assign it to the prototypes $\{p^{m_1}_{c_j}\}_{j\neq i}$ from other clusters $\{c_j\}_{j\neq i}$ as negative pairs. Based on this, we define the 
uni-modal prototypical loss as:

\begin{equation}
\label{eq:uni_modal_loss}
\begin{aligned}
    \mathcal{L}_{\text{Uni}}=-\underset{i\in [K]}{\mathbb{E}}  \left[ \frac{1}{2} \log  \frac{\exp({\cos(r^{m_1}_i, p^{m_1}_{c_i}) /\phi_i^{m_1}})}{ \sum_{j\in [K]} \exp({\cos(r^{m_1}_i, p^{m_1}_{c_j}) /\phi_i^{m_1}}) } + \right. \\
     \left. \frac{1}{2} \log  \frac{\exp({\cos(r^{m_2}_i, p^{m_2}_{c_i}) /\phi_i^{m_2}})}{ \sum_{j\in [K]} \exp({\cos(r^{m_2}_i, p^{m_2}_{c_j}) /\phi_i^{m_2}}) }\right],
\end{aligned}
\end{equation}

\noindent where $\phi$ estimates and balances the concentrations $\phi$ of formed uni-modal clusters. The value of $\phi$ for each cluster is calculated based on the number of samples in this cluster and the average distance between the representations of this cluster and its prototype. The $\phi$ for each modality is then defined as:

\begin{equation}
    \phi^{m}_{i} = \frac{\mathbb{E}_{r^{m}_j \in C^{m}_i}||r^{m}_j - p^{m}_i||_2}{\log(|C^{m}_i| + \alpha)}, m \in \{m_1,m_2\},
\end{equation}

\noindent where $|C^{m}_i|$ is the number of samples in $i^{\text{th}}$ cluster and $\alpha=10$ is a smoothing parameter to prevent producing an overly-large $\phi$. Our final MMSSL objective is:

\begin{equation}
\label{eq.mmssl}
    \mathcal{L} = \frac{1}{2}\mathcal{L}_{\text{Cross}}+\frac{1}{2}\mathcal{L}_{\text{Uni}}.
\end{equation}

\subsection{Select Data with Good Modality Alignment for Cold-start AL}
\label{sec:data_selection}
After obtaining the learned multimodal representations through our designed MMSSL, we need to conduct multimodal data selection to form the initial batch for cold-start AL. We first provide preliminary studies about ActiveFT \cite{activeft}, which is the current state-of-the-art cold-start AL algorithm. As proposed in ActiveFT, the selected subset $X^S$ should have two characteristic: \textbf{(1)} the distribution gap between the representations of the selected subset $R^S$ and those of the entire unlabeled dataset $R^U$ should be minimized and \textbf{(2)} the diversity of the selected subset $R^S$ should be maintained to prevent collapsed selection. 

To align with these two guidelines, the data samples to be selected with the labelling budget $B$ are modeled as a parametric model $\Theta^S = \{\theta_j\}_{j\in[B]}$. In our case with two modalities, we model them as $\Theta^S = \{\theta_j^{m_1} \oplus \theta_j^{m_2}\}_{j\in[B]}$, where each parameter is concatenated by two uni-modal parameters, and it will be a multimodal representation corresponding to one selected multimodal data sample pair. Assume the model is optimized to satisfy all the characteristics, then the sample $x_j$ will be selected if its representation $\langle r^{m_1}_i \oplus r^{m_2}_i \rangle$ is closest to the parameter $\langle \theta^{m_1}_i \oplus \theta^{m_2}_i \rangle$, so that we will have a selected subset accordingly. The optimization goal of the data selection parametric model is:

\begin{equation}
\label{eq.activeft}
    \Theta^S_{\text{opt}} = \arg \underset{\Theta^S}{\min} \mathcal{D}(R^U, \Theta^S) - \lambda_{\text{div}} \mathcal{R}(\Theta^S),
\end{equation}

\noindent where $\mathcal{D}(R^U, \Theta^S)$ is the measurement of distribution gap between the representations of the entire unlabeled dataset and our data selection parametric model, and $\mathcal{R}(\Theta^S)$ is the diversity of the parameters and $\lambda_{\text{div}}$ is a hyper-parameter set as $1.0$ in ActiveFT that controls the contribution of the diversity:

\begin{equation}
\begin{aligned}
\label{eq:disgap}
    \mathcal{D}(R^U, \Theta^S) & = -\underset{r_i\in R^U}{\mathbb{E}}\left[\frac{\cos(r_i^{m_1}, \theta_{s_i}^{m_1}) + \cos(r_i^{m_2}, \theta_{s_i}^{m_2})}{2\tau} \right] , \\
    s_i & =\arg\underset{j\in[B]}{\max}\left[\cos(r_i^{m_1}, \theta_{j}^{m_1}) + \cos(r_i^{m_2}, \theta_{j}^{m_2})\right],
\end{aligned}
\end{equation}

\begin{equation}
\label{eq:diversity}
    \mathcal{R}(\Theta^S) = - \underset{j\in[B]}{\mathbb{E}}\left[\log\sum_{k\neq j}\exp\left(\frac{\cos(\theta_j^{m_1}, \theta_{k}^{m_1})  + \cos(\theta_j^{m_2}, \theta_{k}^{m_2})}{2\tau} \right)\right].
\end{equation}

We argue that, when selecting data pairs with multiple modalities, the selected subset $X^S$ should have an additional characteristic: \textbf{(3)} the cross-modal representation alignment should be maximized within the selected subset for the downstream task model to exploit more shared information across modalities. Therefore we propose our method to improve the modal alignment of the initial batch as shown in the right side of \textbf{Figure \ref{fig:mmssl}}. We add a novel regularization term that specifically maximizes the modality alignment only for the selected subset, independently calculated from the distribution gap term in \textbf{Equation \ref{eq:disgap}} and the diversity term \textbf{Equation \ref{eq:diversity}}. We design this cross-modal alignment regularization term as the average cross-modal similarity between each parameter with all the other parameters in the data selection parametric model:

\begin{equation}
    \mathcal{A}(\Theta^S) = \underset{j\in[B]}{\mathbb{E}}\left[\log\sum_{k\in[B]}\exp\left(\cos(\theta_j^{m_1}, \theta_{k}^{m_2})/\tau \right)\right].
\end{equation}

By including this alignment regularization term with a hyper-parameter $\lambda_{\text{align}}$, controlling the contribution of cross-modal alignment, our final optimization objective is: 

\begin{equation}
\label{eq.mmcsal}
\begin{aligned}
    \mathcal{L} & = \mathcal{D}(R^U,\Theta^S) - \lambda_{\text{div}}\mathcal{R}(\Theta^S) + \lambda_{\text{align}}\mathcal{A}(\Theta^S).
\end{aligned}
\end{equation}

After optimizing the parametric model by minimizing the loss, we select the samples exhibiting the smallest uni-modal distances with the parameters $\Theta^S = \{\theta_j^{m_1} \oplus \theta_j^{m_2}\}_{j\in[B]}$. These samples form our labeled subset $X^S = \{(x^{m_1}_j, x^{m_2}_j)\}_{j\in[B]}$, which is then used to train a multimodal downstream task model. 


\section{Experiments}

\subsection{Dataset}
We choose three different multimodal classification datasets:

\textbf{KineticsSound} \cite{DBLP:conf/iccv/ArandjelovicZ17} is a subset of Kinetics-400 \cite{DBLP:journals/corr/KayCSZHVVGBNSZ17}. It consists of 31 action classes which are all correlated to both video and audio signals. To obtain more data pairs for us to perform MMSSL, we include all available video clips of these 31 action classes from Kinetics-400, resulting in 22,588 video clips for training and 3,012 video clips for testing.

\textbf{Food101} \cite{DBLP:conf/icmcs/WangKTCP15} is an image-text classification dataset for food recipe recognition with 101 kinds of food, where each recipe is composed of a food image and a text recipe description. There are 45,719 sample pairs for training and 15,294 sample pairs for testing.

\textbf{VGGSound} \cite{DBLP:conf/icassp/ChenXVZ20} is a large video-audio dataset with 309 classes where each video clip captures the object that makes the sound. We are only able to download 180,911 clips for training and 14,843 clips for testing. Some of the video clips could not be downloaded due to their unavailability on the YouTube website.

{
\setlength{\tabcolsep}{2.8pt}
\begin{table*}[!htb]
\centering
\fontsize{8.5pt}{10pt}\selectfont  
\begin{tabular}{c|cccc|cccc|cccc}
\toprule
\multirow{2}{*}{AL   Method} & \multicolumn{4}{c|}{Food101}                  & \multicolumn{4}{c|}{KineticsSound}                       & \multicolumn{4}{c}{VGGSound}                  \\
                             & 1\%       & 2\%       & 5\%       & 10\%      & 1\%       & 2\%       & 5\%       & 10\%      & 1\%       & 2\%       & 5\%       & 10\%      \\ \midrule
Random                       & 27.9$\pm$1.5 & 48.1$\pm$1.4 & 67.7$\pm$0.2 & 75.5$\pm$0.1 & 18.9$\pm$1.4 & 26.5$\pm$0.9 & 37.4$\pm$1.3 & 46.5$\pm$1.0 & 14.7$\pm$0.4 & 21.9$\pm$0.3 & 31.7$\pm$0.5 & 38.6$\pm$0.4 \\ \midrule
BADGE                        & -         & 48.4$\pm$1.1 & 68.0$\pm$0.3 & 76.0$\pm$0.3 & -         & 26.2$\pm$0.7 & 36.9$\pm$1.4 & 46.0$\pm$0.8 & -         & 22.1$\pm$0.5 & 32.4$\pm$0.4 & 39.7$\pm$0.2 \\
BMMAL                        & -         & 48.0$\pm$1.8 & 68.1$\pm$0.4 & 76.0$\pm$0.1 & -         & 25.6$\pm$1.3 & 37.2$\pm$0.7 & 46.6$\pm$0.8 & -         & 22.0$\pm$0.2 & 32.9$\pm$0.2 & \textbf{40.0}$\pm$0.3 \\
GCNAL                        & -         & 43.0$\pm$1.1 & 59.5$\pm$2.2 & 69.5$\pm$2.3 & -         & 22.3$\pm$0.8 & 24.9$\pm$1.3 & 32.9$\pm$2.2 & -         & 20.2$\pm$0.4 & 28.9$\pm$0.8 & 35.7$\pm$1.1 \\
ALFA-Mix                     & -         & 46.2$\pm$1.0 & 60.2$\pm$0.8 & 70.0$\pm$0.3 & -         & 26.0$\pm$1.6 & 36.8$\pm$0.6 & 44.9$\pm$3.3 & -         & 20.9$\pm$0.4 & 25.3$\pm$0.3 & 34.5$\pm$0.1 \\ \midrule
KGC                          & 15.3$\pm$1.2 & 26.1$\pm$1.2 & 52.6$\pm$1.0 & 71.2$\pm$0.3 & 14.6$\pm$1.1 & 19.4$\pm$1.5 & 29.0$\pm$1.2 & 38.4$\pm$1.4 & 12.0$\pm$1.4 & 17.4$\pm$0.4 & 28.4$\pm$0.2 & 36.7$\pm$0.1 \\
K-means                      & 23.9$\pm$0.5 & 37.7$\pm$0.5 & 47.9$\pm$0.4 & 59.9$\pm$0.4 & 20.9$\pm$0.6 & 22.9$\pm$0.5 & 32.8$\pm$0.6 & 35.9$\pm$1.0 & 12.1$\pm$0.2 & 15.4$\pm$0.1 & 22.9$\pm$0.3 & 30.1$\pm$0.2 \\
ActiveFT                     & 34.5$\pm$0.9 & 52.6$\pm$1.0 & 68.9$\pm$0.5 & 76.1$\pm$0.2 & 21.9$\pm$0.9 & 28.5$\pm$0.9 & 38.2$\pm$0.7 & 47.0$\pm$0.7 & 18.7$\pm$0.2 & 24.9$\pm$0.4 & 33.1$\pm$0.5 & 39.3$\pm$0.1 \\ \midrule
MMCSAL-proto               & \underline{35.2}$\pm$0.7 & \underline{53.2}$\pm$0.7 & \underline{69.2}$\pm$0.4 & 76.3$\pm$0.4 & \underline{22.4}$\pm$1.6 & \underline{29.0}$\pm$0.6 & \underline{38.8}$\pm$0.6 & 47.2$\pm$0.9 & 18.6$\pm$0.3 & 24.5$\pm$0.3 & 33.4$\pm$0.2 & 39.4$\pm$0.5 \\
MMCSAL-align               & 34.6$\pm$0.2 & 52.7$\pm$0.8 & 68.8$\pm$0.4 & \underline{76.5}$\pm$0.3 & \textbf{22.6}$\pm$1.0 & 28.5$\pm$1.1 & 38.8$\pm$0.8 & \underline{47.4}$\pm$1.1 & \underline{18.8}$\pm$0.4 & \underline{25.1}$\pm$0.4 & \underline{33.9}$\pm$0.4 & \underline{39.8}$\pm$0.3 \\
MMCSAL-final                       & \textbf{36.7}$\pm$1.3 & \textbf{53.7}$\pm$0.3 & \textbf{69.7}$\pm$0.3 & \textbf{76.7}$\pm$0.2 & 22.3$\pm$1.5 & \textbf{29.1}$\pm$0.4 & \textbf{39.7}$\pm$1.0 & \textbf{47.9}$\pm$0.6 & \textbf{19.2}$\pm$0.2 & \textbf{25.6}$\pm$0.2 & \textbf{34.0}$\pm$0.3 & \textbf{40.0}$\pm$0.3 \\ \bottomrule
\end{tabular}
\caption{The supervised cold-start AL experiment results. We report the mean and std of Top-1 accuracy with multiple runs. The best is marked \textbf{bold}, the second best is underlined.}
\label{table:supervised_csal}
\end{table*}
}

\subsection{Baseline} 
We compare our method with random selection, four cold-start AL strategies and four warm-start AL methods. \textbf{Random} selects the data samples randomly from the unlabeled data pool. 

\textbf{Warm-start AL:} \textbf{(1) BADGE} \cite{BADGE} extracts the gradient embedding of multimodal classifier and employs K-means++ initialization method to select diverse and informative data samples for label annotations. \textbf{(2) BMMAL} \cite{BMMAL} achieves balanced multimodal data selection by modulating the gradient embedding of each uni-modal classifier with modality contribution scores. \textbf{(3) GCNAL} \cite{GCNAL} builds a graph convolution network that learns how to classify labeled and unlabeled samples. It chooses those unlabeled samples that are difficult to distinguish. \textbf{(4) ALFA-Mix} \cite{alfa_mix} interpolates the features of unlabeled samples with those of labeled samples, and marks the unlabeled samples as informative ones if their model predictions change after mixing features. The K-means is applied to ensure the diversity of selected subset.

\textbf{Cold-start AL:} \textbf{(1) KGC} (K-greedy-center) \cite{Coreset} greedily selects the sample that has maximum distance from previously selected samples. The Euclidean distance over concatenated multimodal representations is utilized as the distance metric. \textbf{(2)} \textbf{K-means} forms $B$ clusters over concatenated multimodal representations and picks the center sample of each cluster as queried samples. \textbf{(3) ActiveFT} \cite{activeft} models the data selection process as an optimization process that can minimize the distribution gap between the selected subset and the entire unlabeled dataset while maintaining the diversity.

\subsection{Experiment Setting}

\hspace{\parindent} \textbf{Multimodal SSL}: We set the temperature $\tau$ in \textbf{Equations \ref{eq:cross_modal_loss}, \ref{eq:uni_modal_loss}} as $0.07$ for both cross-modal contrastive learning and uni-modal prototypical learning. The batch sizes are all set as 256 for all three datasets. The numbers of uni-modal prototypes are set to 500, 1000, 5000 for KineticsSound, Food101 and VGGSound, respectively, given their different dataset sizes.

\textbf{Data Selection}: We fix the diversity parameter $\lambda_{\text{div}}$ and the temperature $\tau$ in \textbf{Equation \ref{eq.mmcsal}} to $1.0$ and $0.07$, respectively, as in \cite{activeft}. The alignment parameter $\lambda_{\text{align}}$ is set to $1.0$ by default. We examine the impact of this parameter in our supplementary materials.

\textbf{Other Settings:} For Food101, we choose pretrained ResNet-101 \cite{DBLP:conf/cvpr/HeZRS16} as the image backbone and pretrained Bert-base model \cite{DBLP:conf/naacl/DevlinCLT19} as the text backbone. For KineticsSound and VGGSound, we use ResNet2P1D-18 \cite{DBLP:conf/cvpr/TranWTRLP18} as video backbone and ResNet-18 as the audio backbone, modifying the input channel from 3 to 1. We use AdamW \cite{DBLP:conf/iclr/LoshchilovH19} as the optimizer. We repeat 10 runs for Food101 and KineticsSound and 5 runs for VGGSound due to its larger data size.

{
\setlength{\tabcolsep}{2.8pt}
\begin{table*}[!htb]
\centering
\fontsize{8.5pt}{10pt}\selectfont  
\begin{tabular}{c|cccc|cccc|cccc}
\toprule
\multirow{2}{*}{AL   Method} & \multicolumn{4}{c|}{Food101}                  & \multicolumn{4}{c|}{KineticsSound}            & \multicolumn{4}{c}{VGGSound}                  \\
                             & 1\%       & 2\%       & 5\%       & 10\%      & 1\%       & 2\%       & 5\%       & 10\%      & 1\%       & 2\%       & 5\%       & 10\%      \\ \midrule
Random                       & 61.9$\pm$1.6 & 72.2$\pm$0.3 & 77.7$\pm$0.3 & 80.1$\pm$0.1 & 31.6$\pm$2.1 & 41.8$\pm$1.4 & 51.8$\pm$0.9 & 57.1$\pm$0.4 & 29.8$\pm$0.4 & 35.8$\pm$0.3 & 42.2$\pm$0.3 & 45.7$\pm$0.4 \\ \midrule
BADGE \cite{BADGE}                        & -         & 72.7$\pm$0.6 & 78.3$\pm$0.1 & 81.1$\pm$0.2 & -         & 41.7$\pm$1.5 & {52.1}$\pm$0.8 & 57.0$\pm$0.7 & -         & {37.1}$\pm$0.4 & {43.3}$\pm$0.2 & \underline{47.2}$\pm$0.1 \\
BMMAL \cite{BMMAL}                        & -         & 72.4$\pm$0.1 & {78.4}$\pm$0.1 & \underline{81.1$\pm$0.1} & -         & {41.8}$\pm$1.2 & 51.7$\pm$0.6 & {57.0$\pm$0.4} & -         & 36.8$\pm$0.2 & 43.0$\pm$0.3 & {\textbf{47.3}}$\pm$0.2 \\
GCNAL \cite{GCNAL}                        & -         & 67.9$\pm$1.2 & 70.8$\pm$1.1 & 72.9$\pm$0.3 & -         & 36.1$\pm$1.9 & 41.1$\pm$2.1 & 49.1$\pm$2.0 & -         & 34.7$\pm$0.2 & 40.3$\pm$0.2 & 43.9$\pm$0.2 \\
ALFA-Mix \cite{alfa_mix}                     & -         & {73.6}$\pm$0.5 & 78.1$\pm$0.4 & 79.8$\pm$0.5 & -         & 41.8$\pm$1.4 & 52.1$\pm$0.4 & 56.9$\pm$0.6 & -         & 35.2$\pm$0.5 & 40.9$\pm$0.4 & 44.4$\pm$0.5 \\ \midrule
KGC \cite{Coreset}                      & 59.4$\pm$1.2 & 69.3$\pm$0.8 & 77.5$\pm$0.4 & 81.0$\pm$0.2 & 31.6$\pm$0.7 & 40.7$\pm$1.3 & 50.8$\pm$0.9 & 56.8$\pm$0.3 & 31.3$\pm$0.7 & 36.8$\pm$0.4 & 43.0$\pm$0.3 & 46.6$\pm$0.4 \\
K-means                       & 44.5$\pm$0.6 & 62.6$\pm$0.4 & 70.8$\pm$0.1 & 75.1$\pm$0.2 & 25.0$\pm$0.9 & 34.7$\pm$1.1 & 43.3$\pm$0.8 & 48.0$\pm$0.6 & 23.4$\pm$0.1 & 28.0$\pm$0.1 & 36.6$\pm$0.3 & 40.8$\pm$0.3 \\
ActiveFT \cite{activeft}                     & 69.4$\pm$0.6 & \underline{74.2}$\pm$0.8 & 78.5$\pm$0.1 & 80.8$\pm$0.5 & \underline{37.1}$\pm$1.1 & 45.0$\pm$0.7 & 52.9$\pm$0.6 & 57.6$\pm$0.9 & \underline{33.6}$\pm$0.2 & \underline{38.1}$\pm$0.3 & \underline{43.4}$\pm$0.1 & 46.1$\pm$0.3 \\
MMCSAL-final                      & \underline{69.7}$\pm$1.0 & \textbf{74.4}$\pm$0.5 & \textbf{78.9}$\pm$0.2 & \textbf{81.2}$\pm$0.2 & 36.6$\pm$1.8 & \textbf{45.7$\pm$1.1} & \textbf{53.8}$\pm$0.7 & \textbf{57.9}$\pm$0.6 & \textbf{34.3}$\pm$0.3 & \textbf{38.7}$\pm$0.3 & \textbf{43.9}$\pm$0.6 & 47.1$\pm$0.4 \\ 
\bottomrule
\end{tabular}
\caption{The semi-supervised cold-start AL results. We report the mean and std of Top-1 accuracy with multiple runs. The best is marked \textbf{bold}, the second best is underlined.}
\label{table:pretrain-finetune}
\end{table*}
}

\subsection{Supervised Cold-start AL}
\label{sec:supervised_csal}
We examine the performance of different AL strategies in a cold-start setting where only 1\% to 10\% labeling budget is available. \textbf{Table \ref{table:supervised_csal}} shows the performance comparison using \textit{Top-1 Accuracy} as the evaluation metric.

\subsubsection{Warm-start AL Performs Poorly with Insufficient Labels}
The performance of warm-start AL methods with 1\% of labeling data is not shown because they require a random initial subset for model training, which aids in assessing data uncertainty and diversity. As shown in the table, the initially insufficient labeling budget leads to a poorly calibrated model, which fails to guide these warm-start AL methods in selecting informative samples effectively. Consequently, these methods are not as competitive and may even perform worse than random data selection in a cold-start setting with an inadequate label budget.

\subsubsection{Diversity-base Cold-start AL Fails}
KGC and K-means are two diversity-based cold-start AL methods that fail under certain conditions. KGC fails because its greedy selection prioritizes outliers that are difficult for the models to learn, as we provide the evidence in our supplementary materials. K-means fails because it selects only typical data samples, the centroids of clusters, without choosing samples with more uncertainty, thereby providing less novel information. As the table reflects, when the labeling budget increases, the performance gap between K-means and other methods widens due to its tendency to select samples based solely on representativeness, neglecting novel knowledge.

\subsubsection{Uni-modal Prototypes and Alignment Regularization are Compatible and Improve Performance}
\textbf{MMCSAL-proto} only utilizes uni-modal prototypes without the alignment regularization. It performs better than ActiveFT, demonstrating the effectiveness of these prototypes. As discussed in \textbf{Sec \ref{sec:mmssl}} and evidenced in \textbf{Sec \ref{sec:modalit_gap}}, our introduced uni-modal prototypes help bridge the modality gap and balance densities of uni-modal representations, making the representations more organized. Therefore, it estimates the distances between multimodal pairs more precisely, leading to a performance improvement.\textbf{ MMCSAL-align} only adds the regularization term to the data selection process without using uni-modal prototypes. It brings performance gains against ActiveFT by independently enhancing cross-modal alignment within the selected subset and provides more complementary cross-modal information than ActiveFT. We provide additional studies about the alignment parameter $\lambda_{\text{align}}$ in our supplementary materials. Furthermore, \textbf{MMCSAL-final} combines the uni-modal prototypes and the alignment regularization term, achieving the best performance overall. This indicates that these two designed components are compatible with each other and work well together in multimodal cold-start AL.

{
\setlength{\tabcolsep}{2.8pt}
\begin{table}[t!]
\centering
\fontsize{8.5pt}{10pt}\selectfont
\begin{tabular}{c|c|cccc}
\toprule
Datasets & Proto & \multicolumn{1}{c}{$\Delta_{\text{gap}}$} & \multicolumn{1}{c}{$\mathcal{S}^{m_1}_{\text{Uni}}$} & \multicolumn{1}{c}{$\mathcal{S}^{m_2}_{\text{Uni}}$} & $\mathcal{S}_{\text{Cross}}$ \\ \midrule
Food101                        & w/o    & 0.2701   & 0.0204    & 0.0256   & -0.0134      \\
\footnotesize $m_1$:Image\quad $m_2$:Text   & w/      & 0.1203\ $\downarrow$   & 0.0144    & 0.0159   & 0.0079       \\ \midrule
KineticsSound & w/o    & 0.1643   & 0.0188    & 0.0107   & 0.0012       \\
\footnotesize $m_1$:Audio\quad $m_2$:Video   & w/      & 0.0979\ $\downarrow$   & 0.0283    & 0.0201   & 0.0199       \\ \midrule
VGGSound      & w/o    & 0.1327   & 0.0058    & 0.0044   & -0.0037      \\
\footnotesize $m_1$:Audio\quad $m_2$:Video  & w/      & 0.0483\ $\downarrow$   & 0.0073    & 0.0052   & 0.0051       \\ \bottomrule
\end{tabular}
\caption{The modality gap and the expectation of average cosine uni-modal and cross-modal similarity of all samples in three different multimodal datasets. \textit{w/o} represents without prototypes, and \textit{w/} represents with prototypes.}
\label{table0:modality_gap}
\end{table}
}

\subsection{Semi-supervised Cold-start AL}

We conduct semi-supervised learning in cold-start AL where the initially selected labeled samples are used to fine-tune the pretrained models produced with MMSSL. The results are shown in \textbf{Table \ref{table:pretrain-finetune}}. As expected, the semi-supervised learning boosts the downstream task performance by a large margin compared with supervised learning as the model sees more data.  In semi-supervised learning, when labeling budget is 10\%, the warm-start AL methods start to dominate the performance as they can train a well-calibrated model with sufficient data and guide the model to select more informative samples for future AL iterations. However, when available labeling budget is lower than 10\%, warm-start AL methods are not competitive with cold-start AL methods or even close to random data selection. This demonstrates the effectiveness of conducting cold-start AL strategies when labeling information is not sufficient for semi-supervised learning. Overall, our proposed method MMCSAL-final is effective when labeling budget is constrained in semi-supervised learning compared with other cold-start AL strategies, showing that enhancing modality representation and alignment during data selection is a crucial step for multimodal active learning.

\subsection{Modality Gap}
\label{sec:modalit_gap}
To validate that the uni-modal prototypes effectively bridge the modality gap and balance the distance calculations between uni-modal and cross-modal representations, we assess the modality gap \cite{ModalityGap} as the Euclidean distance between the centroid representations of each modality. Additionally, we measure the representation densities by calculating the expectation of the average cosine similarity within and between modalities across the entire dataset:

\begin{equation}
\begin{aligned}
      \mathcal{S}_{\text{Uni}}^{m_k} & = \underset{i\in [N]}{\mathbb{E}} \left[\frac{1}{N}\sum\nolimits_{j \in [N]} \cos(r^{m_k}_i, r^{m_k}_j) \right], k \in \{1, 2\} \\
      \mathcal{S}_{\text{Cross}} & = \underset{i\in [N]}{\mathbb{E}} \left[\frac{1}{N}\sum\nolimits_{j \in [N]} \cos(r^{m_1}_i, r^{m_2}_j) \right] \\
      \Delta_{\text{gap}} & = \left\Vert\frac{1}{N}\sum\nolimits_{i\in[N]} r_i^{m_1} - \frac{1}{N}\sum\nolimits_{i\in[N]}r_i^{m_2}\right\Vert_2 .
\end{aligned}
\end{equation}

These metrics allow us to quantify the effectiveness of the uni-modal prototypes in minimizing the modality gap and enhancing representation similarity assessments. The results are shown in \textbf{Table \ref{table0:modality_gap}}. Multimodal contrastive learning without uni-modal prototypes shows a modality gap between the centroids of the representations of each modality, along with a discrepancy between uni- and cross-modal similarities. With the integration of uni-modal prototypes, we restructure the modality representations, reducing both the modality gap and the disparities between average uni-modal and cross-modal similarities. Consequently, we can more accurately calculate the distances between two multimodal sample pairs, considering both uni-modal and cross-modal distances.

\section{Conclusion}

In this work, we propose a two-stage method to address the cold-start problem in multimodal AL. The introduced uni-modal prototypes bridge the modality gap created by cross-modal contrastive learning. With our MMSSL, the distance estimation among multimodal data pairs becomes more precise, and it benefits data selection. Moreover, we propose to increase the modality alignment for multimodal data pairs to provide more useful modality shared information for downstream multimodal classification tasks. We believe MMCSAL can save labeling budgets for multimodal learning and will explore more multimodal tasks as future work.

\bibliographystyle{ACM-Reference-Format}
\bibliography{reference_v2}


\begin{thebibliography}{39}


\ifx \showCODEN    \undefined \def \showCODEN     #1{\unskip}     \fi
\ifx \showDOI      \undefined \def \showDOI       #1{#1}\fi
\ifx \showISBNx    \undefined \def \showISBNx     #1{\unskip}     \fi
\ifx \showISBNxiii \undefined \def \showISBNxiii  #1{\unskip}     \fi
\ifx \showISSN     \undefined \def \showISSN      #1{\unskip}     \fi
\ifx \showLCCN     \undefined \def \showLCCN      #1{\unskip}     \fi
\ifx \shownote     \undefined \def \shownote      #1{#1}          \fi
\ifx \showarticletitle \undefined \def \showarticletitle #1{#1}   \fi
\ifx \showURL      \undefined \def \showURL       {\relax}        \fi
\providecommand\bibfield[2]{#2}
\providecommand\bibinfo[2]{#2}
\providecommand\natexlab[1]{#1}
\providecommand\showeprint[2][]{arXiv:#2}

\bibitem[Alwassel et~al\mbox{.}(2020)]%
        {DBLP:conf/nips/Alwassel0KTGT20}
\bibfield{author}{\bibinfo{person}{Humam Alwassel}, \bibinfo{person}{Dhruv Mahajan}, \bibinfo{person}{Bruno Korbar}, \bibinfo{person}{Lorenzo Torresani}, \bibinfo{person}{Bernard Ghanem}, {and} \bibinfo{person}{Du Tran}.} \bibinfo{year}{2020}\natexlab{}.
\newblock \showarticletitle{Self-Supervised Learning by Cross-Modal Audio-Video Clustering}. In \bibinfo{booktitle}{\emph{Advances in Neural Information Processing Systems 33: Annual Conference on Neural Information Processing Systems 2020, NeurIPS 2020, December 6-12, 2020, virtual}}.
\newblock
\urldef\tempurl%
\url{https://proceedings.neurips.cc/paper/2020/hash/6f2268bd1d3d3ebaabb04d6b5d099425-Abstract.html}
\showURL{%
\tempurl}


\bibitem[Arandjelovic and Zisserman(2017)]%
        {DBLP:conf/iccv/ArandjelovicZ17}
\bibfield{author}{\bibinfo{person}{Relja Arandjelovic} {and} \bibinfo{person}{Andrew Zisserman}.} \bibinfo{year}{2017}\natexlab{}.
\newblock \showarticletitle{Look, Listen and Learn}. In \bibinfo{booktitle}{\emph{{IEEE} International Conference on Computer Vision, {ICCV} 2017, Venice, Italy, October 22-29, 2017}}. \bibinfo{publisher}{{IEEE} Computer Society}, \bibinfo{pages}{609--617}.
\newblock
\urldef\tempurl%
\url{https://doi.org/10.1109/ICCV.2017.73}
\showDOI{\tempurl}


\bibitem[Ash et~al\mbox{.}(2020)]%
        {BADGE}
\bibfield{author}{\bibinfo{person}{Jordan~T. Ash}, \bibinfo{person}{Chicheng Zhang}, \bibinfo{person}{Akshay Krishnamurthy}, \bibinfo{person}{John Langford}, {and} \bibinfo{person}{Alekh Agarwal}.} \bibinfo{year}{2020}\natexlab{}.
\newblock \showarticletitle{Deep Batch Active Learning by Diverse, Uncertain Gradient Lower Bounds}. In \bibinfo{booktitle}{\emph{8th International Conference on Learning Representations, {ICLR} 2020, Addis Ababa, Ethiopia, April 26-30, 2020}}. \bibinfo{publisher}{OpenReview.net}.
\newblock
\urldef\tempurl%
\url{https://openreview.net/forum?id=ryghZJBKPS}
\showURL{%
\tempurl}


\bibitem[Baltrusaitis et~al\mbox{.}(2019)]%
        {DBLP:journals/pami/BaltrusaitisAM19}
\bibfield{author}{\bibinfo{person}{Tadas Baltrusaitis}, \bibinfo{person}{Chaitanya Ahuja}, {and} \bibinfo{person}{Louis{-}Philippe Morency}.} \bibinfo{year}{2019}\natexlab{}.
\newblock \showarticletitle{Multimodal Machine Learning: {A} Survey and Taxonomy}.
\newblock \bibinfo{journal}{\emph{{IEEE} Trans. Pattern Anal. Mach. Intell.}} \bibinfo{volume}{41}, \bibinfo{number}{2} (\bibinfo{year}{2019}), \bibinfo{pages}{423--443}.
\newblock
\urldef\tempurl%
\url{https://doi.org/10.1109/TPAMI.2018.2798607}
\showDOI{\tempurl}


\bibitem[Caramalau et~al\mbox{.}(2021)]%
        {GCNAL}
\bibfield{author}{\bibinfo{person}{Razvan Caramalau}, \bibinfo{person}{Binod Bhattarai}, {and} \bibinfo{person}{Tae{-}Kyun Kim}.} \bibinfo{year}{2021}\natexlab{}.
\newblock \showarticletitle{Sequential Graph Convolutional Network for Active Learning}. In \bibinfo{booktitle}{\emph{{IEEE} Conference on Computer Vision and Pattern Recognition, {CVPR} 2021, virtual, June 19-25, 2021}}. \bibinfo{publisher}{Computer Vision Foundation / {IEEE}}, \bibinfo{pages}{9583--9592}.
\newblock
\urldef\tempurl%
\url{https://doi.org/10.1109/CVPR46437.2021.00946}
\showDOI{\tempurl}


\bibitem[Caron et~al\mbox{.}(2020)]%
        {DBLP:conf/nips/CaronMMGBJ20}
\bibfield{author}{\bibinfo{person}{Mathilde Caron}, \bibinfo{person}{Ishan Misra}, \bibinfo{person}{Julien Mairal}, \bibinfo{person}{Priya Goyal}, \bibinfo{person}{Piotr Bojanowski}, {and} \bibinfo{person}{Armand Joulin}.} \bibinfo{year}{2020}\natexlab{}.
\newblock \showarticletitle{Unsupervised Learning of Visual Features by Contrasting Cluster Assignments}. In \bibinfo{booktitle}{\emph{Advances in Neural Information Processing Systems 33: Annual Conference on Neural Information Processing Systems 2020, NeurIPS 2020, December 6-12, 2020, virtual}}.
\newblock
\urldef\tempurl%
\url{https://proceedings.neurips.cc/paper/2020/hash/70feb62b69f16e0238f741fab228fec2-Abstract.html}
\showURL{%
\tempurl}


\bibitem[Caron et~al\mbox{.}(2021)]%
        {DBLP:conf/iccv/CaronTMJMBJ21}
\bibfield{author}{\bibinfo{person}{Mathilde Caron}, \bibinfo{person}{Hugo Touvron}, \bibinfo{person}{Ishan Misra}, \bibinfo{person}{Herv{\'{e}} J{\'{e}}gou}, \bibinfo{person}{Julien Mairal}, \bibinfo{person}{Piotr Bojanowski}, {and} \bibinfo{person}{Armand Joulin}.} \bibinfo{year}{2021}\natexlab{}.
\newblock \showarticletitle{Emerging Properties in Self-Supervised Vision Transformers}. In \bibinfo{booktitle}{\emph{2021 {IEEE/CVF} International Conference on Computer Vision, {ICCV} 2021, Montreal, QC, Canada, October 10-17, 2021}}. \bibinfo{publisher}{{IEEE}}, \bibinfo{pages}{9630--9640}.
\newblock
\urldef\tempurl%
\url{https://doi.org/10.1109/ICCV48922.2021.00951}
\showDOI{\tempurl}


\bibitem[Chen et~al\mbox{.}(2020b)]%
        {DBLP:conf/icassp/ChenXVZ20}
\bibfield{author}{\bibinfo{person}{Honglie Chen}, \bibinfo{person}{Weidi Xie}, \bibinfo{person}{Andrea Vedaldi}, {and} \bibinfo{person}{Andrew Zisserman}.} \bibinfo{year}{2020}\natexlab{b}.
\newblock \showarticletitle{Vggsound: {A} Large-Scale Audio-Visual Dataset}. In \bibinfo{booktitle}{\emph{2020 {IEEE} International Conference on Acoustics, Speech and Signal Processing, {ICASSP} 2020, Barcelona, Spain, May 4-8, 2020}}. \bibinfo{publisher}{{IEEE}}, \bibinfo{pages}{721--725}.
\newblock
\urldef\tempurl%
\url{https://doi.org/10.1109/ICASSP40776.2020.9053174}
\showDOI{\tempurl}


\bibitem[Chen et~al\mbox{.}(2023)]%
        {chen2024making}
\bibfield{author}{\bibinfo{person}{Liangyu Chen}, \bibinfo{person}{Yutong Bai}, \bibinfo{person}{Siyu Huang}, \bibinfo{person}{Yongyi Lu}, \bibinfo{person}{Bihan Wen}, \bibinfo{person}{Alan~L. Yuille}, {and} \bibinfo{person}{Zongwei Zhou}.} \bibinfo{year}{2023}\natexlab{}.
\newblock \showarticletitle{Making Your First Choice: To Address Cold Start Problem in Medical Active Learning}. In \bibinfo{booktitle}{\emph{Medical Imaging with Deep Learning, {MIDL} 2023, 10-12 July 2023, Nashville, TN, {USA}}} \emph{(\bibinfo{series}{Proceedings of Machine Learning Research}, Vol.~\bibinfo{volume}{227})}. \bibinfo{publisher}{{PMLR}}, \bibinfo{pages}{496--525}.
\newblock
\urldef\tempurl%
\url{https://proceedings.mlr.press/v227/chen24a.html}
\showURL{%
\tempurl}


\bibitem[Chen et~al\mbox{.}(2020a)]%
        {mocov2}
\bibfield{author}{\bibinfo{person}{Xinlei Chen}, \bibinfo{person}{Haoqi Fan}, \bibinfo{person}{Ross~B. Girshick}, {and} \bibinfo{person}{Kaiming He}.} \bibinfo{year}{2020}\natexlab{a}.
\newblock \showarticletitle{Improved Baselines with Momentum Contrastive Learning}.
\newblock \bibinfo{journal}{\emph{CoRR}}  \bibinfo{volume}{abs/2003.04297} (\bibinfo{year}{2020}).
\newblock
\showeprint[arXiv]{2003.04297}
\urldef\tempurl%
\url{https://arxiv.org/abs/2003.04297}
\showURL{%
\tempurl}


\bibitem[Devlin et~al\mbox{.}(2019)]%
        {DBLP:conf/naacl/DevlinCLT19}
\bibfield{author}{\bibinfo{person}{Jacob Devlin}, \bibinfo{person}{Ming{-}Wei Chang}, \bibinfo{person}{Kenton Lee}, {and} \bibinfo{person}{Kristina Toutanova}.} \bibinfo{year}{2019}\natexlab{}.
\newblock \showarticletitle{{BERT:} Pre-training of Deep Bidirectional Transformers for Language Understanding}. In \bibinfo{booktitle}{\emph{Proceedings of the 2019 Conference of the North American Chapter of the Association for Computational Linguistics: Human Language Technologies, {NAACL-HLT} 2019, Minneapolis, MN, USA, June 2-7, 2019, Volume 1 (Long and Short Papers)}}, \bibfield{editor}{\bibinfo{person}{Jill Burstein}, \bibinfo{person}{Christy Doran}, {and} \bibinfo{person}{Thamar Solorio}} (Eds.). \bibinfo{publisher}{Association for Computational Linguistics}, \bibinfo{pages}{4171--4186}.
\newblock
\urldef\tempurl%
\url{https://doi.org/10.18653/V1/N19-1423}
\showDOI{\tempurl}


\bibitem[Elizalde et~al\mbox{.}(2023)]%
        {clap}
\bibfield{author}{\bibinfo{person}{Benjamin Elizalde}, \bibinfo{person}{Soham Deshmukh}, \bibinfo{person}{Mahmoud~Al Ismail}, {and} \bibinfo{person}{Huaming Wang}.} \bibinfo{year}{2023}\natexlab{}.
\newblock \showarticletitle{{CLAP} Learning Audio Concepts from Natural Language Supervision}. In \bibinfo{booktitle}{\emph{{IEEE} International Conference on Acoustics, Speech and Signal Processing {ICASSP} 2023, Rhodes Island, Greece, June 4-10, 2023}}. \bibinfo{publisher}{{IEEE}}, \bibinfo{pages}{1--5}.
\newblock
\urldef\tempurl%
\url{https://doi.org/10.1109/ICASSP49357.2023.10095889}
\showDOI{\tempurl}


\bibitem[Hacohen et~al\mbox{.}(2022)]%
        {al_on_a_budget}
\bibfield{author}{\bibinfo{person}{Guy Hacohen}, \bibinfo{person}{Avihu Dekel}, {and} \bibinfo{person}{Daphna Weinshall}.} \bibinfo{year}{2022}\natexlab{}.
\newblock \showarticletitle{Active Learning on a Budget: Opposite Strategies Suit High and Low Budgets}. In \bibinfo{booktitle}{\emph{International Conference on Machine Learning, {ICML} 2022, 17-23 July 2022, Baltimore, Maryland, {USA}}} \emph{(\bibinfo{series}{Proceedings of Machine Learning Research}, Vol.~\bibinfo{volume}{162})}. \bibinfo{publisher}{{PMLR}}, \bibinfo{pages}{8175--8195}.
\newblock
\urldef\tempurl%
\url{https://proceedings.mlr.press/v162/hacohen22a.html}
\showURL{%
\tempurl}


\bibitem[He et~al\mbox{.}(2020)]%
        {DBLP:conf/cvpr/He0WXG20}
\bibfield{author}{\bibinfo{person}{Kaiming He}, \bibinfo{person}{Haoqi Fan}, \bibinfo{person}{Yuxin Wu}, \bibinfo{person}{Saining Xie}, {and} \bibinfo{person}{Ross~B. Girshick}.} \bibinfo{year}{2020}\natexlab{}.
\newblock \showarticletitle{Momentum Contrast for Unsupervised Visual Representation Learning}. In \bibinfo{booktitle}{\emph{2020 {IEEE/CVF} Conference on Computer Vision and Pattern Recognition, {CVPR} 2020, Seattle, WA, USA, June 13-19, 2020}}. \bibinfo{publisher}{Computer Vision Foundation / {IEEE}}, \bibinfo{pages}{9726--9735}.
\newblock
\urldef\tempurl%
\url{https://doi.org/10.1109/CVPR42600.2020.00975}
\showDOI{\tempurl}


\bibitem[He et~al\mbox{.}(2016)]%
        {DBLP:conf/cvpr/HeZRS16}
\bibfield{author}{\bibinfo{person}{Kaiming He}, \bibinfo{person}{Xiangyu Zhang}, \bibinfo{person}{Shaoqing Ren}, {and} \bibinfo{person}{Jian Sun}.} \bibinfo{year}{2016}\natexlab{}.
\newblock \showarticletitle{Deep Residual Learning for Image Recognition}. In \bibinfo{booktitle}{\emph{2016 {IEEE} Conference on Computer Vision and Pattern Recognition, {CVPR} 2016, Las Vegas, NV, USA, June 27-30, 2016}}. \bibinfo{publisher}{{IEEE} Computer Society}, \bibinfo{pages}{770--778}.
\newblock
\urldef\tempurl%
\url{https://doi.org/10.1109/CVPR.2016.90}
\showDOI{\tempurl}


\bibitem[Jenni et~al\mbox{.}(2023)]%
        {DBLP:conf/aaai/Jenni0C23}
\bibfield{author}{\bibinfo{person}{Simon Jenni}, \bibinfo{person}{Alexander Black}, {and} \bibinfo{person}{John~P. Collomosse}.} \bibinfo{year}{2023}\natexlab{}.
\newblock \showarticletitle{Audio-Visual Contrastive Learning with Temporal Self-Supervision}. In \bibinfo{booktitle}{\emph{Thirty-Seventh {AAAI} Conference on Artificial Intelligence, {AAAI} 2023, Thirty-Fifth Conference on Innovative Applications of Artificial Intelligence, {IAAI} 2023, Thirteenth Symposium on Educational Advances in Artificial Intelligence, {EAAI} 2023, Washington, DC, USA, February 7-14, 2023}}. \bibinfo{publisher}{{AAAI} Press}, \bibinfo{pages}{7996--8004}.
\newblock
\urldef\tempurl%
\url{https://doi.org/10.1609/AAAI.V37I7.25967}
\showDOI{\tempurl}


\bibitem[Johnson et~al\mbox{.}(2019)]%
        {johnson2019billion}
\bibfield{author}{\bibinfo{person}{Jeff Johnson}, \bibinfo{person}{Matthijs Douze}, {and} \bibinfo{person}{Herv{\'e} J{\'e}gou}.} \bibinfo{year}{2019}\natexlab{}.
\newblock \showarticletitle{Billion-scale similarity search with {GPUs}}.
\newblock \bibinfo{journal}{\emph{IEEE Transactions on Big Data}} \bibinfo{volume}{7}, \bibinfo{number}{3} (\bibinfo{year}{2019}), \bibinfo{pages}{535--547}.
\newblock


\bibitem[Kay et~al\mbox{.}(2017)]%
        {DBLP:journals/corr/KayCSZHVVGBNSZ17}
\bibfield{author}{\bibinfo{person}{Will Kay}, \bibinfo{person}{Jo{\~{a}}o Carreira}, \bibinfo{person}{Karen Simonyan}, \bibinfo{person}{Brian Zhang}, \bibinfo{person}{Chloe Hillier}, \bibinfo{person}{Sudheendra Vijayanarasimhan}, \bibinfo{person}{Fabio Viola}, \bibinfo{person}{Tim Green}, \bibinfo{person}{Trevor Back}, \bibinfo{person}{Paul Natsev}, \bibinfo{person}{Mustafa Suleyman}, {and} \bibinfo{person}{Andrew Zisserman}.} \bibinfo{year}{2017}\natexlab{}.
\newblock \showarticletitle{The Kinetics Human Action Video Dataset}.
\newblock \bibinfo{journal}{\emph{CoRR}}  \bibinfo{volume}{abs/1705.06950} (\bibinfo{year}{2017}).
\newblock
\showeprint[arXiv]{1705.06950}
\urldef\tempurl%
\url{http://arxiv.org/abs/1705.06950}
\showURL{%
\tempurl}


\bibitem[Li et~al\mbox{.}(2021)]%
        {prototypical}
\bibfield{author}{\bibinfo{person}{Junnan Li}, \bibinfo{person}{Pan Zhou}, \bibinfo{person}{Caiming Xiong}, {and} \bibinfo{person}{Steven C.~H. Hoi}.} \bibinfo{year}{2021}\natexlab{}.
\newblock \showarticletitle{Prototypical Contrastive Learning of Unsupervised Representations}. In \bibinfo{booktitle}{\emph{9th International Conference on Learning Representations, {ICLR} 2021, Virtual Event, Austria, May 3-7, 2021}}. \bibinfo{publisher}{OpenReview.net}.
\newblock
\urldef\tempurl%
\url{https://openreview.net/forum?id=KmykpuSrjcq}
\showURL{%
\tempurl}


\bibitem[Liang et~al\mbox{.}(2022)]%
        {ModalityGap}
\bibfield{author}{\bibinfo{person}{Weixin Liang}, \bibinfo{person}{Yuhui Zhang}, \bibinfo{person}{Yongchan Kwon}, \bibinfo{person}{Serena Yeung}, {and} \bibinfo{person}{James~Y. Zou}.} \bibinfo{year}{2022}\natexlab{}.
\newblock \showarticletitle{Mind the Gap: Understanding the Modality Gap in Multi-modal Contrastive Representation Learning}. In \bibinfo{booktitle}{\emph{Advances in Neural Information Processing Systems 35: Annual Conference on Neural Information Processing Systems 2022, NeurIPS 2022, New Orleans, LA, USA, November 28 - December 9, 2022}}.
\newblock
\urldef\tempurl%
\url{http://papers.nips.cc/paper\_files/paper/2022/hash/702f4db7543a7432431df588d57bc7c9-Abstract-Conference.html}
\showURL{%
\tempurl}


\bibitem[Loshchilov and Hutter(2019)]%
        {DBLP:conf/iclr/LoshchilovH19}
\bibfield{author}{\bibinfo{person}{Ilya Loshchilov} {and} \bibinfo{person}{Frank Hutter}.} \bibinfo{year}{2019}\natexlab{}.
\newblock \showarticletitle{Decoupled Weight Decay Regularization}. In \bibinfo{booktitle}{\emph{7th International Conference on Learning Representations, {ICLR} 2019, New Orleans, LA, USA, May 6-9, 2019}}. \bibinfo{publisher}{OpenReview.net}.
\newblock
\urldef\tempurl%
\url{https://openreview.net/forum?id=Bkg6RiCqY7}
\showURL{%
\tempurl}


\bibitem[Ma et~al\mbox{.}(2021)]%
        {DBLP:conf/iclr/MaZMS21}
\bibfield{author}{\bibinfo{person}{Shuang Ma}, \bibinfo{person}{Zhaoyang Zeng}, \bibinfo{person}{Daniel McDuff}, {and} \bibinfo{person}{Yale Song}.} \bibinfo{year}{2021}\natexlab{}.
\newblock \showarticletitle{Active Contrastive Learning of Audio-Visual Video Representations}. In \bibinfo{booktitle}{\emph{9th International Conference on Learning Representations, {ICLR} 2021, Virtual Event, Austria, May 3-7, 2021}}. \bibinfo{publisher}{OpenReview.net}.
\newblock
\urldef\tempurl%
\url{https://openreview.net/forum?id=OMizHuea\_HB}
\showURL{%
\tempurl}


\bibitem[Parvaneh et~al\mbox{.}(2022)]%
        {alfa_mix}
\bibfield{author}{\bibinfo{person}{Amin Parvaneh}, \bibinfo{person}{Ehsan Abbasnejad}, \bibinfo{person}{Damien Teney}, \bibinfo{person}{Reza Haffari}, \bibinfo{person}{Anton van~den Hengel}, {and} \bibinfo{person}{Javen~Qinfeng Shi}.} \bibinfo{year}{2022}\natexlab{}.
\newblock \showarticletitle{Active Learning by Feature Mixing}. In \bibinfo{booktitle}{\emph{{IEEE/CVF} Conference on Computer Vision and Pattern Recognition, {CVPR} 2022, New Orleans, LA, USA, June 18-24, 2022}}. \bibinfo{publisher}{{IEEE}}, \bibinfo{pages}{12227--12236}.
\newblock
\urldef\tempurl%
\url{https://doi.org/10.1109/CVPR52688.2022.01192}
\showDOI{\tempurl}


\bibitem[Radford et~al\mbox{.}(2021)]%
        {clip}
\bibfield{author}{\bibinfo{person}{Alec Radford}, \bibinfo{person}{Jong~Wook Kim}, \bibinfo{person}{Chris Hallacy}, \bibinfo{person}{Aditya Ramesh}, \bibinfo{person}{Gabriel Goh}, \bibinfo{person}{Sandhini Agarwal}, \bibinfo{person}{Girish Sastry}, \bibinfo{person}{Amanda Askell}, \bibinfo{person}{Pamela Mishkin}, \bibinfo{person}{Jack Clark}, \bibinfo{person}{Gretchen Krueger}, {and} \bibinfo{person}{Ilya Sutskever}.} \bibinfo{year}{2021}\natexlab{}.
\newblock \showarticletitle{Learning Transferable Visual Models From Natural Language Supervision}. In \bibinfo{booktitle}{\emph{Proceedings of the 38th International Conference on Machine Learning, {ICML} 2021, 18-24 July 2021, Virtual Event}} \emph{(\bibinfo{series}{Proceedings of Machine Learning Research}, Vol.~\bibinfo{volume}{139})}. \bibinfo{publisher}{{PMLR}}, \bibinfo{pages}{8748--8763}.
\newblock
\urldef\tempurl%
\url{http://proceedings.mlr.press/v139/radford21a.html}
\showURL{%
\tempurl}


\bibitem[Ren et~al\mbox{.}(2022)]%
        {DBLP:journals/csur/RenXCHLGCW22}
\bibfield{author}{\bibinfo{person}{Pengzhen Ren}, \bibinfo{person}{Yun Xiao}, \bibinfo{person}{Xiaojun Chang}, \bibinfo{person}{Po{-}Yao Huang}, \bibinfo{person}{Zhihui Li}, \bibinfo{person}{Brij~B. Gupta}, \bibinfo{person}{Xiaojiang Chen}, {and} \bibinfo{person}{Xin Wang}.} \bibinfo{year}{2022}\natexlab{}.
\newblock \showarticletitle{A Survey of Deep Active Learning}.
\newblock \bibinfo{journal}{\emph{{ACM} Comput. Surv.}} \bibinfo{volume}{54}, \bibinfo{number}{9} (\bibinfo{year}{2022}), \bibinfo{pages}{180:1--180:40}.
\newblock
\urldef\tempurl%
\url{https://doi.org/10.1145/3472291}
\showDOI{\tempurl}


\bibitem[Sener and Savarese(2018)]%
        {Coreset}
\bibfield{author}{\bibinfo{person}{Ozan Sener} {and} \bibinfo{person}{Silvio Savarese}.} \bibinfo{year}{2018}\natexlab{}.
\newblock \showarticletitle{Active Learning for Convolutional Neural Networks: {A} Core-Set Approach}. In \bibinfo{booktitle}{\emph{6th International Conference on Learning Representations, {ICLR} 2018, Vancouver, BC, Canada, April 30 - May 3, 2018, Conference Track Proceedings}}. \bibinfo{publisher}{OpenReview.net}.
\newblock
\urldef\tempurl%
\url{https://openreview.net/forum?id=H1aIuk-RW}
\showURL{%
\tempurl}


\bibitem[Shen et~al\mbox{.}(2023)]%
        {BMMAL}
\bibfield{author}{\bibinfo{person}{Meng Shen}, \bibinfo{person}{Yizheng Huang}, \bibinfo{person}{Jianxiong Yin}, \bibinfo{person}{Heqing Zou}, \bibinfo{person}{Deepu Rajan}, {and} \bibinfo{person}{Simon See}.} \bibinfo{year}{2023}\natexlab{}.
\newblock \showarticletitle{Towards Balanced Active Learning for Multimodal Classification}. In \bibinfo{booktitle}{\emph{Proceedings of the 31st {ACM} International Conference on Multimedia, {MM} 2023, Ottawa, ON, Canada, 29 October 2023- 3 November 2023}}. \bibinfo{publisher}{{ACM}}, \bibinfo{pages}{3434--3445}.
\newblock
\urldef\tempurl%
\url{https://doi.org/10.1145/3581783.3612463}
\showDOI{\tempurl}


\bibitem[Shi et~al\mbox{.}(2022)]%
        {DBLP:conf/iclr/ShiHLM22}
\bibfield{author}{\bibinfo{person}{Bowen Shi}, \bibinfo{person}{Wei{-}Ning Hsu}, \bibinfo{person}{Kushal Lakhotia}, {and} \bibinfo{person}{Abdelrahman Mohamed}.} \bibinfo{year}{2022}\natexlab{}.
\newblock \showarticletitle{Learning Audio-Visual Speech Representation by Masked Multimodal Cluster Prediction}. In \bibinfo{booktitle}{\emph{The Tenth International Conference on Learning Representations, {ICLR} 2022, Virtual Event, April 25-29, 2022}}. \bibinfo{publisher}{OpenReview.net}.
\newblock
\urldef\tempurl%
\url{https://openreview.net/forum?id=Z1Qlm11uOM}
\showURL{%
\tempurl}


\bibitem[Swayamdipta et~al\mbox{.}(2020)]%
        {dataset-map}
\bibfield{author}{\bibinfo{person}{Swabha Swayamdipta}, \bibinfo{person}{Roy Schwartz}, \bibinfo{person}{Nicholas Lourie}, \bibinfo{person}{Yizhong Wang}, \bibinfo{person}{Hannaneh Hajishirzi}, \bibinfo{person}{Noah~A. Smith}, {and} \bibinfo{person}{Yejin Choi}.} \bibinfo{year}{2020}\natexlab{}.
\newblock \showarticletitle{Dataset Cartography: Mapping and Diagnosing Datasets with Training Dynamics}. In \bibinfo{booktitle}{\emph{Proceedings of the 2020 Conference on Empirical Methods in Natural Language Processing, {EMNLP} 2020, Online, November 16-20, 2020}}, \bibfield{editor}{\bibinfo{person}{Bonnie Webber}, \bibinfo{person}{Trevor Cohn}, \bibinfo{person}{Yulan He}, {and} \bibinfo{person}{Yang Liu}} (Eds.). \bibinfo{publisher}{Association for Computational Linguistics}, \bibinfo{pages}{9275--9293}.
\newblock
\urldef\tempurl%
\url{https://doi.org/10.18653/V1/2020.EMNLP-MAIN.746}
\showDOI{\tempurl}


\bibitem[Tran et~al\mbox{.}(2018)]%
        {DBLP:conf/cvpr/TranWTRLP18}
\bibfield{author}{\bibinfo{person}{Du Tran}, \bibinfo{person}{Heng Wang}, \bibinfo{person}{Lorenzo Torresani}, \bibinfo{person}{Jamie Ray}, \bibinfo{person}{Yann LeCun}, {and} \bibinfo{person}{Manohar Paluri}.} \bibinfo{year}{2018}\natexlab{}.
\newblock \showarticletitle{A Closer Look at Spatiotemporal Convolutions for Action Recognition}. In \bibinfo{booktitle}{\emph{2018 {IEEE} Conference on Computer Vision and Pattern Recognition, {CVPR} 2018, Salt Lake City, UT, USA, June 18-22, 2018}}. \bibinfo{publisher}{Computer Vision Foundation / {IEEE} Computer Society}, \bibinfo{pages}{6450--6459}.
\newblock
\urldef\tempurl%
\url{https://doi.org/10.1109/CVPR.2018.00675}
\showDOI{\tempurl}


\bibitem[van~den Oord et~al\mbox{.}(2018)]%
        {infonce}
\bibfield{author}{\bibinfo{person}{A{\"{a}}ron van~den Oord}, \bibinfo{person}{Yazhe Li}, {and} \bibinfo{person}{Oriol Vinyals}.} \bibinfo{year}{2018}\natexlab{}.
\newblock \showarticletitle{Representation Learning with Contrastive Predictive Coding}.
\newblock \bibinfo{journal}{\emph{CoRR}}  \bibinfo{volume}{abs/1807.03748} (\bibinfo{year}{2018}).
\newblock
\showeprint[arXiv]{1807.03748}
\urldef\tempurl%
\url{http://arxiv.org/abs/1807.03748}
\showURL{%
\tempurl}


\bibitem[Wang et~al\mbox{.}(2015)]%
        {DBLP:conf/icmcs/WangKTCP15}
\bibfield{author}{\bibinfo{person}{Xin Wang}, \bibinfo{person}{Devinder Kumar}, \bibinfo{person}{Nicolas Thome}, \bibinfo{person}{Matthieu Cord}, {and} \bibinfo{person}{Fr{\'{e}}d{\'{e}}ric Precioso}.} \bibinfo{year}{2015}\natexlab{}.
\newblock \showarticletitle{Recipe recognition with large multimodal food dataset}. In \bibinfo{booktitle}{\emph{2015 {IEEE} International Conference on Multimedia {\&} Expo Workshops, {ICME} Workshops 2015, Turin, Italy, June 29 - July 3, 2015}}. \bibinfo{publisher}{{IEEE} Computer Society}, \bibinfo{pages}{1--6}.
\newblock
\urldef\tempurl%
\url{https://doi.org/10.1109/ICMEW.2015.7169757}
\showDOI{\tempurl}


\bibitem[Xie et~al\mbox{.}(2023)]%
        {activeft}
\bibfield{author}{\bibinfo{person}{Yichen Xie}, \bibinfo{person}{Han Lu}, \bibinfo{person}{Junchi Yan}, \bibinfo{person}{Xiaokang Yang}, \bibinfo{person}{Masayoshi Tomizuka}, {and} \bibinfo{person}{Wei Zhan}.} \bibinfo{year}{2023}\natexlab{}.
\newblock \showarticletitle{Active Finetuning: Exploiting Annotation Budget in the Pretraining-Finetuning Paradigm}. In \bibinfo{booktitle}{\emph{{IEEE/CVF} Conference on Computer Vision and Pattern Recognition, {CVPR} 2023, Vancouver, BC, Canada, June 17-24, 2023}}. \bibinfo{publisher}{{IEEE}}, \bibinfo{pages}{23715--23724}.
\newblock
\urldef\tempurl%
\url{https://doi.org/10.1109/CVPR52729.2023.02271}
\showDOI{\tempurl}


\bibitem[Xu et~al\mbox{.}(2021)]%
        {video_clip}
\bibfield{author}{\bibinfo{person}{Hu Xu}, \bibinfo{person}{Gargi Ghosh}, \bibinfo{person}{Po{-}Yao Huang}, \bibinfo{person}{Dmytro Okhonko}, \bibinfo{person}{Armen Aghajanyan}, \bibinfo{person}{Florian Metze}, \bibinfo{person}{Luke Zettlemoyer}, {and} \bibinfo{person}{Christoph Feichtenhofer}.} \bibinfo{year}{2021}\natexlab{}.
\newblock \showarticletitle{VideoCLIP: Contrastive Pre-training for Zero-shot Video-Text Understanding}. In \bibinfo{booktitle}{\emph{Proceedings of the 2021 Conference on Empirical Methods in Natural Language Processing, {EMNLP} 2021, Virtual Event / Punta Cana, Dominican Republic, 7-11 November, 2021}}. \bibinfo{publisher}{Association for Computational Linguistics}, \bibinfo{pages}{6787--6800}.
\newblock
\urldef\tempurl%
\url{https://doi.org/10.18653/V1/2021.EMNLP-MAIN.544}
\showDOI{\tempurl}


\bibitem[Yi et~al\mbox{.}(2022)]%
        {pt4al}
\bibfield{author}{\bibinfo{person}{John Seon~Keun Yi}, \bibinfo{person}{Minseok Seo}, \bibinfo{person}{Jongchan Park}, {and} \bibinfo{person}{Dong{-}Geol Choi}.} \bibinfo{year}{2022}\natexlab{}.
\newblock \showarticletitle{{PT4AL:} Using Self-supervised Pretext Tasks for Active Learning}. In \bibinfo{booktitle}{\emph{Computer Vision - {ECCV} 2022 - 17th European Conference, Tel Aviv, Israel, October 23-27, 2022, Proceedings, Part {XXVI}}} \emph{(\bibinfo{series}{Lecture Notes in Computer Science}, Vol.~\bibinfo{volume}{13686})}. \bibinfo{publisher}{Springer}, \bibinfo{pages}{596--612}.
\newblock
\urldef\tempurl%
\url{https://doi.org/10.1007/978-3-031-19809-0\_34}
\showDOI{\tempurl}


\bibitem[Yuan et~al\mbox{.}(2020)]%
        {cold_start_al_nlp}
\bibfield{author}{\bibinfo{person}{Michelle Yuan}, \bibinfo{person}{Hsuan{-}Tien Lin}, {and} \bibinfo{person}{Jordan~L. Boyd{-}Graber}.} \bibinfo{year}{2020}\natexlab{}.
\newblock \showarticletitle{Cold-start Active Learning through Self-supervised Language Modeling}. In \bibinfo{booktitle}{\emph{Proceedings of the 2020 Conference on Empirical Methods in Natural Language Processing, {EMNLP} 2020, Online, November 16-20, 2020}}. \bibinfo{publisher}{Association for Computational Linguistics}, \bibinfo{pages}{7935--7948}.
\newblock
\urldef\tempurl%
\url{https://doi.org/10.18653/V1/2020.EMNLP-MAIN.637}
\showDOI{\tempurl}


\bibitem[Zhang et~al\mbox{.}(2020)]%
        {DBLP:journals/jstsp/ZhangYHD20}
\bibfield{author}{\bibinfo{person}{Chao Zhang}, \bibinfo{person}{Zichao Yang}, \bibinfo{person}{Xiaodong He}, {and} \bibinfo{person}{Li Deng}.} \bibinfo{year}{2020}\natexlab{}.
\newblock \showarticletitle{Multimodal Intelligence: Representation Learning, Information Fusion, and Applications}.
\newblock \bibinfo{journal}{\emph{{IEEE} J. Sel. Top. Signal Process.}} \bibinfo{volume}{14}, \bibinfo{number}{3} (\bibinfo{year}{2020}), \bibinfo{pages}{478--493}.
\newblock
\urldef\tempurl%
\url{https://doi.org/10.1109/JSTSP.2020.2987728}
\showDOI{\tempurl}


\bibitem[Zolfaghari et~al\mbox{.}(2021)]%
        {crossclr}
\bibfield{author}{\bibinfo{person}{Mohammadreza Zolfaghari}, \bibinfo{person}{Yi Zhu}, \bibinfo{person}{Peter~V. Gehler}, {and} \bibinfo{person}{Thomas Brox}.} \bibinfo{year}{2021}\natexlab{}.
\newblock \showarticletitle{CrossCLR: Cross-modal Contrastive Learning For Multi-modal Video Representations}. In \bibinfo{booktitle}{\emph{2021 {IEEE/CVF} International Conference on Computer Vision, {ICCV} 2021, Montreal, QC, Canada, October 10-17, 2021}}. \bibinfo{publisher}{{IEEE}}, \bibinfo{pages}{1430--1439}.
\newblock
\urldef\tempurl%
\url{https://doi.org/10.1109/ICCV48922.2021.00148}
\showDOI{\tempurl}


\bibitem[Zong et~al\mbox{.}(2023)]%
        {mmssl_survey}
\bibfield{author}{\bibinfo{person}{Yongshuo Zong}, \bibinfo{person}{Oisin~Mac Aodha}, {and} \bibinfo{person}{Timothy~M. Hospedales}.} \bibinfo{year}{2023}\natexlab{}.
\newblock \showarticletitle{Self-Supervised Multimodal Learning: {A} Survey}.
\newblock \bibinfo{journal}{\emph{CoRR}}  \bibinfo{volume}{abs/2304.01008} (\bibinfo{year}{2023}).
\newblock
\urldef\tempurl%
\url{https://doi.org/10.48550/ARXIV.2304.01008}
\showDOI{\tempurl}
\showeprint[arXiv]{2304.01008}


\end{thebibliography}

\clearpage

\section{Implementation Details}
In this section, we provide more details about our implementation of multimodal self-supervised learning and both supervised and semi-supervised cold-start active learning to facilitate reproduction of our reported results.

\subsection{Multimodal Self-supervised Learning}
We perform multimodal self-supervised learning with a local batch size of 64 on each GPU card and a global batch size of 256 across four GPU cards. The unimodal representations are vectors with 512 dimensions. The multimodal representations are concatenated unimodal representations. The learning schedulers are cosine schedulers with linear warm-up. Note that the unimodal prototypical loss is calculated only after the warm-up period. We use one linear layer as a projector to map the outputs from the feature backbones into the unified 512-dimensional representations.

For data augmentation and loading, we follow the settings in BMMAL\cite{BMMAL}. For Food101, we apply image data augmentations such as random resize crop, random horizontal flip, grayscale, and color jittering during training. The cropped images with a size of 224 are used as visual inputs. We use Bert's tokenizer to tokenize the textual food recipes and use the tokens as textual inputs. For KineticsSound and VGGSound, we first uniformly sample 10 frames from each video clip. We then randomly sample 3 frames and apply random resize crop and random horizontal flip as video data augmentations. The cropped 3 video frames with a size of 168 are used as video inputs. We resample audio clips to 16 kHz and randomly select 5-second long audio clips for audio data augmentation. We use the spectrum of the audio as audio inputs. The number of Fourier transform points for each frame in the audio is set to 512, the window size is set to 512, and the hop length is set to 159.

For Food101, we train for 25 epochs and warm up for 5 epochs. The learning rates are set to $1 \times 10^{-5}$ for the text backbone, $1 \times 10^{-4}$ for the image backbone, and $1 \times 10^{-3}$ for projectors. For KineticsSound and VGGSound, we train for 45 epochs and warm up for 15 epochs. The learning rates are set to $1 \times 10^{-3}$ for the video backbone, the audio backbone and the projectors. We use the same AdamW optimizer with a weight decay of 0.02 and betas of 0.9 and 0.95 across the three datasets.

\subsection{Supervised Cold-start AL}

We perform the same data augmentation as in MMSSL. For Food101, we train for 15 epochs and warm up for 5 epochs. The learning rates are set to $1 \times 10^{-5}$ for the text backbone, $1 \times 10^{-4}$ for the image backbone, and $1 \times 10^{-3}$ for the projectors. For video-audio datasets, KineticsSound and VGGSound, we train for 45 epochs and warm up for 15 epochs. The learning rates are set to $1 \times 10^{-3}$ for the video backbone, the audio backbone and the projectors. We use AdamW optimizer with a weight decay of 0.01 and betas of 0.9 and 0.999 for supervised training.

\subsection{Semi-supervised Cold-start AL}

We reduce the number of training epochs and the learning rates to preserve the knowledge learned from the pretraining stage. For Food101, we train for 10 epochs and warm up for 5 epochs. The learning rates are set to $2 \times 10^{-6}$ for the text backbone, $2 \times 10^{-5}$ for the image backbone, and $1 \times 10^{-3}$ for projectors. For KineticsSound, we train for 20 epochs and warm up for 10 epochs. The learning rates are set to $5 \times 10^{-4}$ for the video backbone and the audio backbone and $1 \times 10^{-3}$ for projectors. For VGGSound, we train for 20 epochs and warm up for 10 epochs. The learning rates are set to $1 \times 10^{-4}$ for the video backbone and the audio backbone and $1 \times 10^{-3}$ for projectors. We use the same optimizer as in supervised cold-start active learning.

\subsection{Optimize the Parametric Selection Model}

Following the implementation of ActiveFT \cite{activeft}, we use Adam as the optimizer to optimize the parametric selection model with learning objective in \textbf{Equation 9}. To make optimization easier, parameters are initialized using the representations of uniformly selected samples rather than being randomly initialized. We set the learning rate to $1 \times 10^{-3}$ and use a cosine scheduler to adjust the rate. The model is optimized over 300 epochs. After optimization, we select the samples that have the smallest unimodal distances to the parameters as our labeled subset.

\subsection{Computational Cost}

We present the computational cost for stage 1 and 2 in \textbf{Table \ref{table:stage_1}} and \textbf{Table \ref{table:stage_2}} across three different multimodal datasets. In stage 1, GPU memory consumption is higher due to using FAISS \cite{johnson2019billion}, which requires 1.5 - 2 GB of memory to accelerate k-means with the GPU. Additionally, the prototypes do not consume a significant amount of memory, as shown in \textbf{Table \ref{table:stage_1}}. For the training process with prototypes in stage 1 and the data selection process in stage 2, our method takes on average 20\%-40\% more time, which we find acceptable.

\begin{table}[!t]
\centering
\fontsize{8.5pt}{10pt}\selectfont 
\begin{tabular}{c|c|ccc}
\toprule
                                                                              & Proto & Foo101 & KineticsSound & VGGSound \\ \midrule
\multirow{2}{*}{\begin{tabular}[c]{@{}c@{}}Training Time\\ (Hours)\end{tabular}} & w/o       & 0.6    & 2.6           & 20.7     \\
                                                                              & w/        & 0.9    & 3.5           & 27.5     \\ \midrule
\multirow{2}{*}{\begin{tabular}[c]{@{}c@{}}GPU Memory\\ (GB)\end{tabular}}    & w/o       & 17.2   & 19.9          & 19.9     \\
                                                                              & w/        & 18.8   & 22.1          & 22.2       \\ \bottomrule
\end{tabular}
\caption{The training time and GPU memory for MMSSL with and without prototypes at stage 1.}
\label{table:stage_1}
\end{table}

\begin{table}[!t]
\centering
\fontsize{8.5pt}{10pt}\selectfont 
\begin{tabular}{c|cc|cc|cc}
\toprule
     & \multicolumn{2}{c|}{Foo101} & \multicolumn{2}{c|}{KineticsSound} & \multicolumn{2}{c}{VGGSound} \\ 
     & ActiveFT      & Ours      & ActiveFT          & Ours         & ActiveFT       & Ours      \\ \midrule
1\%  & 9.9           & 11.4        & 7.2              & 9.1           & 65.3          & 74.5      \\ 
2\%  & 12.7          & 14.7        & 7.5              & 9.5           & 111.7         & 131.3      \\ 
5\%  & 21.5          & 27.5        & 9.6              & 12.2          & 255.6         & 342.4      \\ 
10\% & 37.4          & 52.9        & 13.6             & 18.6          & 509.2         & 770.3      \\ \bottomrule
\end{tabular}
\caption{The data selection time in seconds at stage 2.}
\label{table:stage_2}
\end{table}

\section{Analysis of $\lambda_{\text{align}}$}
\label{sec:analysis}
To study the impact of the hyper-parameter $\lambda_{\text{align}}$ in \textbf{Equation 9}, which controls the contribution of cross-modal alignment in the selected subset, we vary this parameter in our supervised cold-start AL experiments. It should be noted that we always keep $\lambda_{\text{div}}$ as $1.0$ which aligns with ActiveFT \cite{activeft} to introduce diversity enhancement. When $\lambda_{\text{align}}$ equals to $0.0$, the method becomes MMCSAL-proto, which only includes diversity enhancement without any cross-modal alignment, and when  $\lambda_{\text{align}}$ equals to $1.0$, it is our proposed method MMCSAL-final, which balances  alignment and diversity. As shown in \textbf{Tabel \ref{table2:ablation_study_lambda_2}}, when $\lambda_{\text{align}}$ is increased from $0.0$ to $1.0$, the supervised AL performance increases, expect in situations where the labeling budget is extremely low for KineticsSound and VGGSound. However, if we keep increasing $\lambda_{\text{align}}$ up to $2.0$, the supervised AL performance decreases. This shows that enhancing cross-modal alignment for the selected subset improves multimodal data quality for downstream multimodal tasks. While diversity may be the priority when the labeling budget is extremely low, such as when the labeling budget is 1 \% or 2 \% for KineticsSound and 1 \% for VGGSound, the better option for $\lambda_{\text{align}}$ should be less than 1.0. Therefore, one may consider using a smaller $\lambda_{\text{align}}$ to balance diversity. Overall, we recommend using $1.0$ as the default value for $\lambda_{\text{align}}$.

{
\setlength{\tabcolsep}{4pt}
\begin{table}[!t]
\centering
\fontsize{8.2pt}{10pt}\selectfont
\begin{tabular}{cccccc}
\toprule
\multicolumn{1}{c|}{Labeling}     & \multicolumn{5}{c}{\textbf{$\lambda_{\text{align}}$}}                                                                                                                                                                                             \\
\multicolumn{1}{c|}{Budget}     & \multicolumn{1}{c}{0.0}      & \multicolumn{1}{c}{0.25} & \multicolumn{1}{c}{0.50} & \multicolumn{1}{c}{1.0}    & 2.0                   \\ \midrule
\multicolumn{1}{c|}{}          & \multicolumn{5}{c}{\textbf{Food101}} \\ 
\multicolumn{1}{c|}{1\%}  & 35.2$\pm$0.7                                                                   & 36.2$\pm$0.7                 & 36.0$\pm$1.0                 & \textbf{36.7}$\pm$1.3                                                                 & 36.6$\pm$1.4 \\
\multicolumn{1}{c|}{2\%}  & 53.2$\pm$0.7                                                                   & 53.3$\pm$1.2                 & 53.6$\pm$0.6                 & \textbf{53.7}$\pm$0.3                                                                 & 53.2$\pm$0.6 \\
\multicolumn{1}{c|}{5\%}  & 69.2$\pm$0.4                                                                   & 69.3$\pm$0.4                 & 69.4$\pm$0.1                 & \textbf{69.7}$\pm$0.3                                                                 & 69.4$\pm$0.2 \\
\multicolumn{1}{c|}{10\%} & 76.3$\pm$0.4                                                                   & 76.5$\pm$0.2                 & 76.4$\pm$0.2                 & \textbf{76.7}$\pm$0.2                                                                 & 75.6$\pm$0.3 \\ \midrule
\multicolumn{1}{c|}{}          & \multicolumn{5}{c}{\textbf{KineticsSound}} \\ 
\multicolumn{1}{c|}{1\%}  & 22.4$\pm$1.6                                                                   & \textbf{23.0}$\pm$1.2                 & 22.7$\pm$1.3                 & 22.3$\pm$1.5                                                                 & 22.1$\pm$1.1 \\
\multicolumn{1}{c|}{2\%}  & 29.0$\pm$0.6                                                                   & 28.7$\pm$0.9                 & \textbf{29.5}$\pm$0.7                 & 29.1$\pm$0.4                                                                & 28.7$\pm$0.9 \\
\multicolumn{1}{c|}{5\%}  & 38.8$\pm$0.6                                                                   & 38.6$\pm$1.1                 & 38.5$\pm$0.7                 & \textbf{39.7}$\pm$1.0                                                                 & 38.9$\pm$0.6 \\
\multicolumn{1}{c|}{10\%} & 47.2$\pm$0.9                                                                   & 47.1$\pm$0.6                 & 47.8$\pm$0.3                 & \textbf{47.9}$\pm$0.6                                                                  & 47.7$\pm$0.6 \\ \midrule
\multicolumn{1}{c|}{}          & \multicolumn{5}{c}{\textbf{VGGSound}} \\ 
\multicolumn{1}{c|}{1\%}  & 18.6$\pm$0.3                                                                   & 19.2$\pm$0.3                 & \textbf{19.3}$\pm$0.2                 & 19.2$\pm$0.2                                                                 & 18.7$\pm$0.2 \\
\multicolumn{1}{c|}{2\%}  & 24.5$\pm$0.3                                                                   & 25.4$\pm$0.3                 & 25.4$\pm$0.5                 & \textbf{25.6}$\pm$0.2                                                                 & 24.9$\pm$0.2 \\
\multicolumn{1}{c|}{5\%}  & 33.4$\pm$0.2                                                                   & 33.9$\pm$0.2                 & 33.6$\pm$0.3                 & \textbf{34.0}$\pm$0.3                                                                 & 33.5$\pm$0.3 \\
\multicolumn{1}{c|}{10\%} & 39.4$\pm$0.5                                                                   & 39.5$\pm$0.2                 & 40.0$\pm$0.4                 & \textbf{40.0$\pm$0.3}                                                                & 38.4$\pm$0.3 \\
\bottomrule
\end{tabular}
\caption{The analysis of $\lambda_{\text{align}}$.}
\label{table2:ablation_study_lambda_2}
\end{table}
}

\section{Data Selection Preference of AL}

\begin{figure}[ht]
  \centering
   \includegraphics[width=0.90\linewidth]{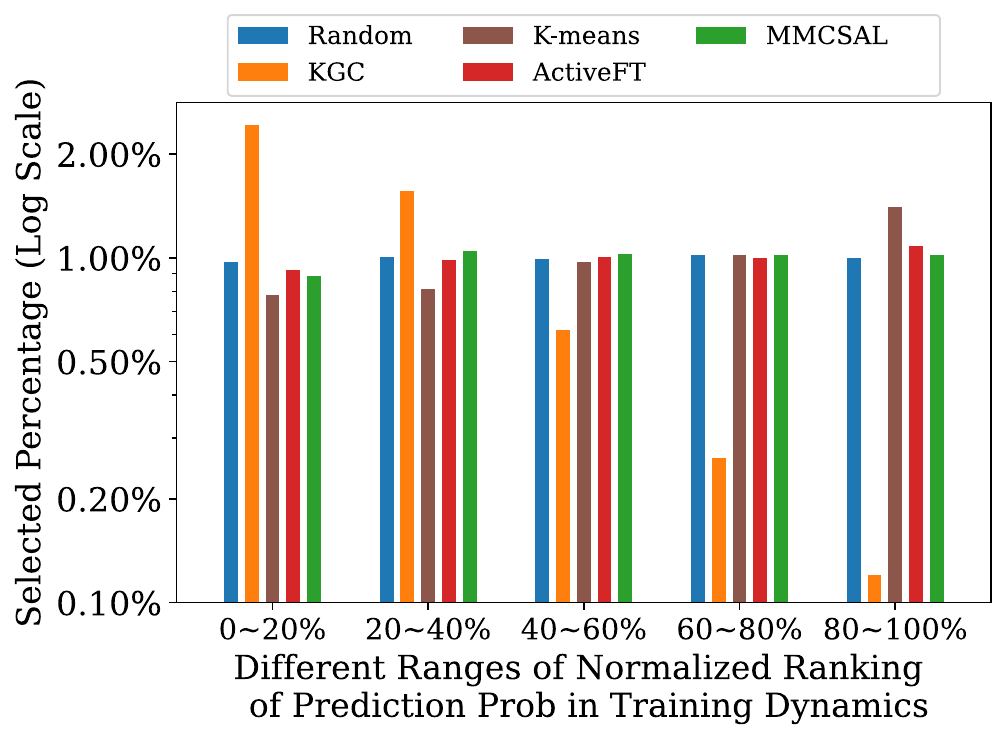}
   \caption{Preference for data selection of different AL strategies with 5\% labeling budget on Food101.}
   \label{fig:food101_data_preference}
\end{figure}

\begin{figure}[ht]
  \centering
   \includegraphics[width=0.9\linewidth]{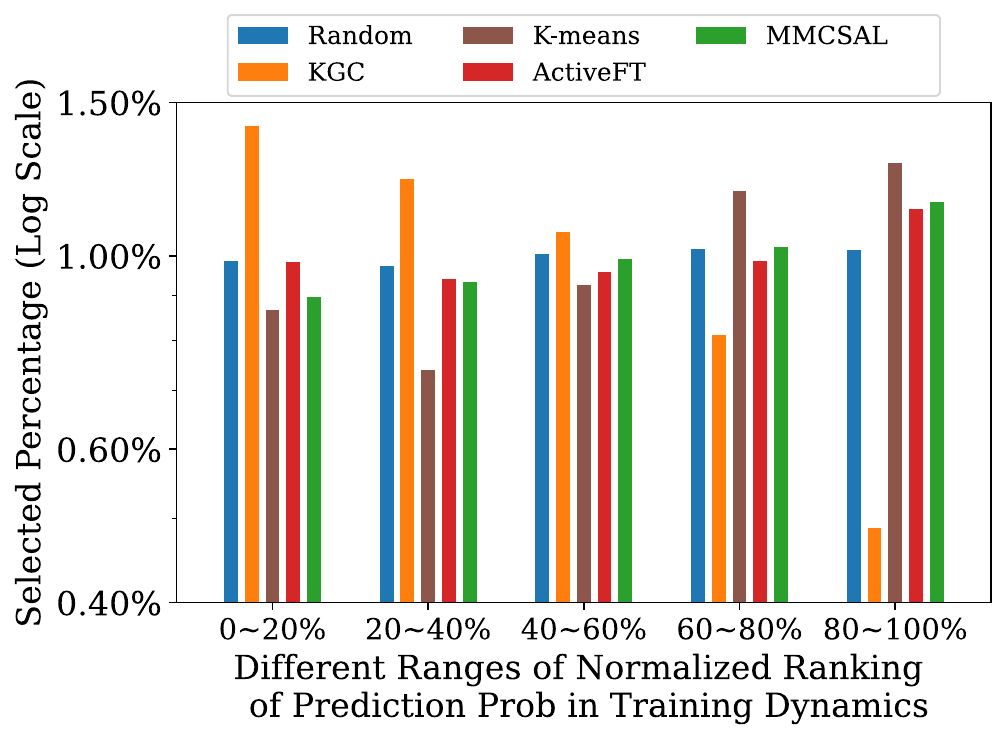}
   \caption{Preference for data selection of different AL strategies with 5\% labeling budget on KineticsSound.}
   \label{fig:kineticssound_data_preference}
\end{figure}

\begin{figure}[ht]
  \centering
   \includegraphics[width=0.9\linewidth]{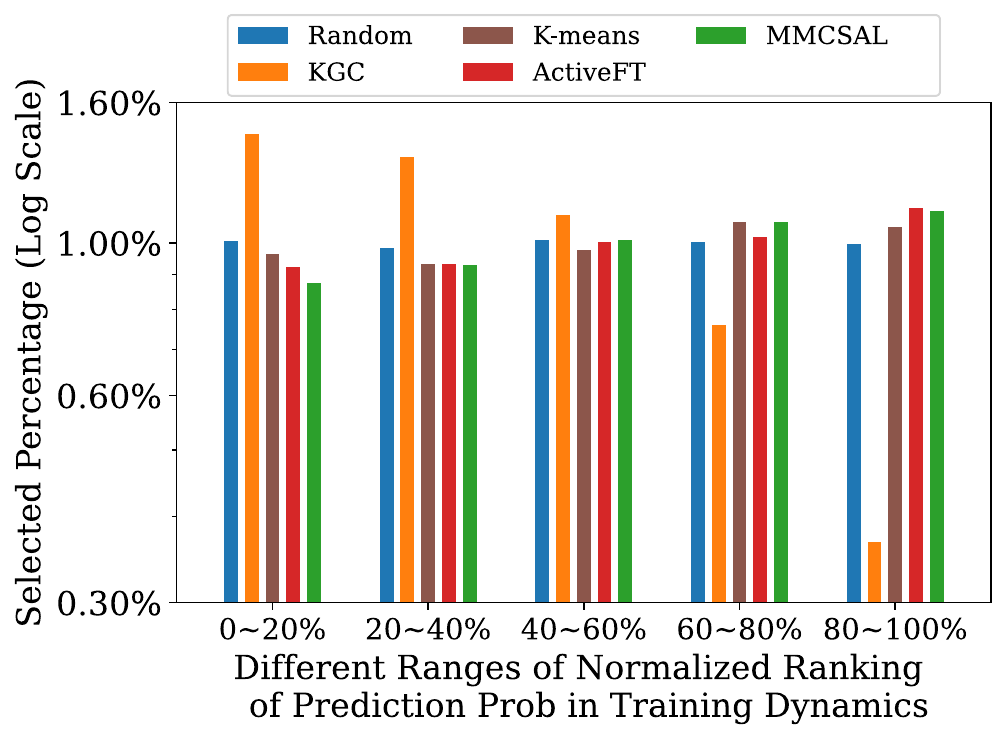}
   \caption{Preference for data selection of different AL strategies with 5\% labeling budget on VGGSound.}
   \label{fig:vggsound_data_preference}
\end{figure}

We first explain how we derive the results of data preference in \textbf{Figures \ref{fig:food101_data_preference}, \ref{fig:kineticssound_data_preference}, \ref{fig:vggsound_data_preference}}. We then present results on the data selection preferences of cold-start active learning strategies across three datasets, Food101, KineticsSound and VGGSound. 

\subsection{Rank the Samples by Confidence}
We perform supervised multimodal classification training with the entire labeled dataset as in \cite{dataset-map}. We calculate the confidence score as the average prediction probability across all training epochs:

\begin{equation}
    \mu_i = \frac{1}{T}\sum_{t=1}^{T}p(y_i|x_i),
\end{equation}

where $T$ is the number of epochs and $p(y_i|x_i)$ is the prediction probability for the ground truth label $y_i$. We then rank all data samples in ascending order and use the normalized ranking scores $r_i$ to indicate their confidence levels:

\begin{equation}
    r_i = \frac{\text{rank}(\mu_i)}{N},
\end{equation}

where $N$ is the number of all samples. A sample with a lower $r_i$ is considered uncertain, while a sample with a higher $r_i$ is considered confident. We group all samples into five groups with different ranges of normalized ranking scores. The bars in the data selection preference figures represent the numbers of selected samples by each cold-start active learning strategy, thereby showing the selection preference for uncertain and confident samples by different active learning strategies.

\subsection{Analyze the Data Preference of AL}

We visualize the preference for data selection of cold-start AL strategies by counting the numbers of confident, ambiguous and uncertain samples in selected subset. The results are presented in \textbf{Figures \ref{fig:food101_data_preference}, \ref{fig:kineticssound_data_preference}, \ref{fig:vggsound_data_preference}}. All the samples are ranked by their average prediction probability over the ground truth class in an ascending order as described above. This probability is calculated from the training dynamics of supervised multimodal classification \cite{dataset-map}. A high-ranking sample is a confident one, while a low-ranking one is an uncertain one. The samples in the middle are the ambiguous samples. As shown in the figures, KGC selects more samples that are uncertain due to its greedy data selection strategy. This explains its failure in cold-start active learning. K-means selects more samples that are confident given the fact that it selects the centroids of clusters which are often typical and easy-to-learn. Excepted for VGGSound, K-means selects more uniformly due to its larger dataset size. Compared with ActiveFT, MMCSAL tends to select less samples that are the most and the least confident ones. The most confident samples are believed to be perfectly aligned but they contribute less on diversity as they are also more similar to other samples. The least confident samples are mostly poorly aligned and they contribute less on modality alignment. Since MMCSAL optimizes both the diversity and the modality alignment within the selected subset, these ambiguous samples may contribute more than these most and least confident samples. It shows that MMCSAL achieves a more balanced data selection between diversity and modality alignment for multimodal data, leading to better cold-start AL performance.

\end{document}